\documentclass[11pt, onecolumn]{IEEEtran}

\usepackage{tabularx}
\usepackage{cancel}
\usepackage{multirow}
\usepackage{supertabular}
\usepackage{algorithmic}
\usepackage{algorithm}
\usepackage{amsthm}
\usepackage{float}
\usepackage{subfig}
\usepackage{rotating}
\usepackage{rotating}
\usepackage{amssymb}
\usepackage{color}
\usepackage{graphicx}
\usepackage{amsmath}
\usepackage{wasysym}
\usepackage{multirow}
\usepackage{booktabs}
\usepackage{lineno}
\usepackage{soul}
\usepackage{authblk}
\usepackage{longtable}
\usepackage{supertabular}

\newcommand{\delOx}{$\delta{}^{18}\mathrm{O}$}
\newcommand{\delOxb}{$\delta{}^{17}\mathrm{O}$}
\newcommand{\delD}{$\delta\mathrm{D}$}
\newcommand{\Dxs}{$\mathrm{d_{xs}}$}
\newcommand{\delN}{$\delta{}^{15}\mathrm{N}$}

\newcommand{\delAr}{$\delta{}^{40}\mathrm{Ar}$}
\newcommand{\delcorrel}{$r_{\delta{}^{18}\mathrm{O}/\delta\mathrm{D}}$}

\begin{document}

\title{Molecular diffusion of stable water isotopes in polar firn as a proxy for past temperatures}

\author[1]{Christian~Holme}
\author[1]{Vasileios~Gkinis}
\author[1]{Bo~M.~Vinther}

\affil[1]{Centre for Ice and Climate, Niels Bohr Institute, University of Copenhagen,
Juliane Maries Vej 30, DK-2100 Copenhagen, Denmark}


\maketitle

\begin{abstract}
Polar precipitation archived in ice caps contains information on past temperature conditions.
Such information can be retrieved by measuring the water isotopic signals of \delOx~and \delD~in ice cores.
These signals have been attenuated during densification due to molecular diffusion in the firn column,
where the magnitude of the diffusion is isotopologoue specific and temperature dependent.
By utilizing the differential diffusion signal, dual isotope measurements of \delOx~and \delD~enable multiple
temperature reconstruction techniques. This study assesses how well six different methods can be used to
reconstruct past surface temperatures from the diffusion-based temperature proxies. Two of the methods are
based on the single diffusion lengths of \delOx~and \delD, three of the methods employ the differential diffusion signal,
while the last uses the ratio between the single diffusion lengths. All techniques are tested on synthetic data
in order to evaluate their accuracy and precision.
We perform a benchmark test to thirteen high resolution Holocene data sets
from Greenland and Antarctica, which represent a broad range of mean annual surface temperatures and accumulation rates.
Based on the benchmark test, we comment on the accuracy and precision of the methods.
Both the benchmark test and the synthetic data test demonstrate
that the most precise reconstructions are obtained when using the single isotope diffusion lengths, with precisions of approximately $1.0\,^\mathrm{o}\mathrm{C}$.
In the benchmark test, the single isotope diffusion lengths are also found to reconstruct consistent temperatures with a
root-mean-square-deviation of $0.7\,^\mathrm{o}\mathrm{C}$.
The techniques employing the differential diffusion signals are more uncertain,
where the most precise method has a precision of $1.9\,^\mathrm{o}\mathrm{C}$.
The diffusion length ratio method is the least precise with a precision of $13.7\,^\mathrm{o}\mathrm{C}$.
The absolute temperature estimates from this method are also shown to be highly sensitive to the choice of fractionation factor parameterization.

\end{abstract}




\section{Introduction} 
\label{Intro}
Polar precipitation stored for thousands of years in the ice caps of Greenland and Antarctica
contains unique information on past climatic conditions. The isotopic composition
of polar ice, commonly expressed
through the $\delta$ notation
has been used as a direct proxy of the relative depletion of a water vapor mass in its
journey from the evaporation site to the place where condensation takes place \cite{Epstein1951, IAEA}.
Additionally, for modern times, the isotopic signal of present day shows a good correlation with the temperature of the cloud at the
time of precipitation \cite{Dansgaard2, Dansgaard} and as a result it has been proposed and used
as a proxy of past temperatures \cite{Jouzel1984, Jouzel1997, Johnsen2001}.

The use of the isotopic paleothermometer presents some notable limitations.
The modern day linear relationship between \delOx~and temperature commonly referred to as the ``spatial slope''
may hold for present conditions,
but studies based on borehole temperature reconstruction \cite{Cuffey1994, JOHNSEN1995a}
as well as the thermal fractionation of the \delN~signal in polar firn
\cite{Severinghaus1998, Severinghaus1999} have
independently underlined the inaccuracy of the spatial isotope slope when it is extrapolated to past climatic
conditions.
Even though qualitatively the \delOx~signal comprises past temperature information, it fails
to provide a quantitative picture on the magnitudes of past climatic changes.

\cite{Johnsen1977, WhillansGrootes1985} and \cite{Johnsen2000} set the foundations for the quantitative description
of the diffusive processes the water isotopic signal undergoes in the porous firn layer from the time
of deposition until pore close--off. Even though the main purpose of \cite{Johnsen2000} was to investigate
how to reconstruct the part of the signal that was attenuated during the diffusive processes, the authors
make a reference to the possibility of using the assessment of the diffusive rates as a proxy for
past firn temperatures.

The temperature reconstruction method based on isotope firn diffusion requires
data of high resolution. Moreover, if one would like to look into the differential diffusion signal,
datasets of both \delOx~and \delD~are required. Such data sets have until recently not been easy to
obtain especially due to the challenging nature of the \delD~analysis \cite{BIGELEISEN1952, Vaughn1998}.
With the advent of present 	commercial high--accuracy, high--precision Infra-Red spectrometers
\cite{Crosson2008, Brand2009}, simultaneous measurements of \delOx~and \delD~have become
easier to obtain. Coupling of these instruments to Continuous Flow Analysis systems \cite{Gkinis2011, Maselli2013, Emanuelsson2015, Jones2017a}
can also result in measurements of ultra--high resolution, a necessary condition for
accurate temperature reconstructions based on water isotope diffusion.

A number of existing works have presented past firn temperature reconstructions
based on water isotope diffusion. \cite{Simonsen2011} and \cite{Gkinis2014} used high resolution
isotopic datasets from the NorthGRIP ice core \cite{NGRIPmembers2004}. The first study makes use
of the differential diffusion signal, utilizing spectral estimates of high--resolution
dual  \delOx~and \delD~datasets covering
the GS--1 and GI--1 periods in the NorthGRIP ice core \cite{Rasmussen2014}.
The second study
presents a combined temperature and accumulation history of the past 16,000 years based on the
power spectral density (\textbf{PSD} hereafter) signals of
high resolution \delOx~measurements of the NorthGRIP ice core. More recently, \cite{vanderWel2015a}
introduced a slightly different approach for reconstructing the differential diffusion signal and testing it
on dual \delOx, \delD~high resolution data from the EDML ice core \cite{Oerter2004}.
By artificially forward--diffusing the \delD~signal the authors estimate differential diffusion rates
by maximizing the correlation between the \delOx~and \delD~signal.
In this work we attempt to test the various approaches in utilizing the temperature
reconstruction technique.

We use synthetic, as well as real ice core data sets that represent
Holocene conditions from a variety of drilling sites on Greenland and Antarctica.
Our objective is to use data sections that originate from parts of the core as close to present day
as possible. By doing this we aim to minimize possible uncertainties and biases in the ice flow thinning
adjustment that is required for temperature interpretation of the diffusion rate estimates.
Such a bias has been shown to exist for the NorthGRIP ice core \cite{Gkinis2014}, most likely
due to the \cite{DJmodel} ice flow model overestimating  the past accumulation rates for the site.
In order to include as much data as possible, approximately half of the datasets used here
have an age of 9-10 ka. This age coincides with the timing of the early Holocene Climate Optimum around 5-9 ka (\textbf{HCO} hereafter).
For Greenlandic drill sites, temperatures were up to $3 \,^\mathrm{o}$C warmer than present day during the HCO \cite{Dorthe1998}.
Another aspect of this study is that it uses water isotopic data sets of \delOx~and
\delD~measured using different analytical techniques, namely Isotope Ratio Mass Spectroscopy
(\textbf{IRMS} hereafter) as  well as Cavity Ring Down Spectroscopy (\textbf{CRDS} hereafter).
Two of the data sets presented here were obtained using Continuous Flow Analysis (\textbf{CFA} hereafter)
systems tailored for water isotopic analysis \cite{Gkinis2011}. All data sections are characterized by
a very high sampling resolution typically of $5\, \mathrm{cm}$ or better.

\section{Theory}

\subsection{Diffusion of water isotope signals in firn}
The porous medium of the top $60-80\, \mathrm{m}$ of firn allows for a molecular diffusion process that attenuates
the water isotope signal from the time of deposition until  pore close--off.
The process takes place in the vapor phase and it can be described by Fick's second law as
(assuming that the water isotope ratio signal ($\delta$) approximates the concentration):
\begin{equation}
\frac{\partial \delta}{\partial t} = D \left( t \right) \frac{\partial^2 \delta}{\partial z^2} -
\dot{\varepsilon}_z \left( t \right) z ~\frac{\partial \delta}{\partial z} \enspace
\label{eq.diffusion}
\end{equation}
where $D \left( t \right)$ is the diffusivity coefficient, $\dot{\varepsilon}_z\left(t\right)$
the vertical strain rate and  $z$ is the vertical axis of a coordinate system,
with its origin being fixed within the considered layer.
The attenuation of the isotopic signal results in loss of information.
However, the dependence of $\dot{\varepsilon}_z\left(t\right)$  and $D \left( t \right)$
on temperature and accumulation presents the possibility of using the process as a tool
to infer these two paleoclimatic parameters. A solution to Eq. \ref{eq.diffusion} can be given by
the convolution of the initial isotopic profile $\delta'$ with a Gaussian filter $\mathcal{G}$ as:
\begin{equation}
\delta \left( z\right) = \mathcal{S} \left( z \right) \left[ \delta ' \left( z \right)
\ast \mathcal{G} \left( z \right) \right]
\label{eq.convolution}
\end{equation}
where the Gaussian filter is described as:
\begin{equation}
\mathcal{G}\left(z \right) = \frac{1}{\sigma \sqrt{2\pi}} \, e^{\frac{-z^2}{2 \sigma^2}} \enspace ,
\label{eq.gaussian}
\end{equation}
and $\mathcal{S}$ is the total thinning of the layer at depth $z$ described by
\begin{equation}
\mathcal{S} \left( z \right) = e^{\int_0^{z} \dot{\varepsilon}_z \left( z' \right) \mathrm{d}z'} \enspace .
\label{eq.thinning}
\end{equation}
In Eq. \ref{eq.gaussian}, the standard deviation term $\sigma^2$ represents the average
displacement of a water molecule along the z--axis and is commonly referred to as the diffusion length.
The $\sigma^2$ quantity is a direct measure of diffusion and its accurate estimate is critical
to any attempt of reconstructing temperatures that are based on the isotope diffusion thermometer.
The diffusion length is directly related to the diffusivity coefficient and the strain rate
(as the strain rate is approximately proportional to the densification rate in the firn column)
and it can therefore be regarded as a measure of firn temperature.

The differential equation describing the evolution of $\sigma^2$ with time is
given by \cite{Johnsen1977}:
\begin{equation}
\frac{\mathrm{d}\sigma^2}{\mathrm{d}t} - 2\,\dot{\varepsilon}_z\!\left( t \right) \sigma^2 =
2D\!\left( t \right) \enspace.
\label{eq.diflength}
\end{equation}
In the case of firn the following approximation can be made for the strain rate:
\begin{equation}
\dot{\varepsilon}_z \left(
	t \right) \approx -\frac{\mathrm{d\rho}}{\,\,\,\mathrm{d t}}\,\frac{\,1\,}{\,\rho\,},
\label{eq.strainrate}
\end{equation}
with $\rho$ representing the density.
Then for the firn column, Eq. \ref{eq.diflength} can be solved hereby yielding a solution for $\sigma^2$:
\begin{equation}
\sigma^2 \left( \rho \right) = \frac{\,1\,}{\rho^2}\int_{\rho_o}^{\rho}2\rho^2
{\left( \frac{\mathrm{d}\rho}{\mathrm{d}t}\right)}^{-1}\! D \!\left( \rho \right) \,\mathrm{d}\rho,
\label{eq.difflengthintegration}
\end{equation}
where  $\rho_o$ is the surface density.
Under the assumption that the diffusivity coefficient $D\left(\rho\right)$ and the
densification rate   $\frac{\mathrm{d}\rho}{\mathrm{d}t}$ are known,
integration from surface density $\rho_o$ to the close--off density $\rho_{co}$
can be performed yielding a model based estimate for the diffusion length.
In this work we make use of the Herron--Langway densification model (\textbf{H--L hereafter})
and the diffusivity rate parametrization introduced by \cite{Johnsen2000} (Supplementary Online Material (\textbf{SOM}) Sec. S1).
$\frac{\mathrm{d}\rho}{\mathrm{d}t}$ depends on temperature and overburden pressure
and $D\left(\rho\right)$ depends on temperature and firn connectivity.
Our implementation of Eq. \ref{eq.difflengthintegration} includes a seasonal temperature signal that propagates down in the firn (SOM Sec. S2).
The seasonal temperature variation affects the firn diffusion length nonlinearly in the upper $10-12 \, \mathrm{m}$ due to the saturation vapor pressure's exponential
dependence on temperature.

\begin{figure}[]
\vspace*{2mm}
	\begin{center}
			\includegraphics[width=0.7\textwidth]{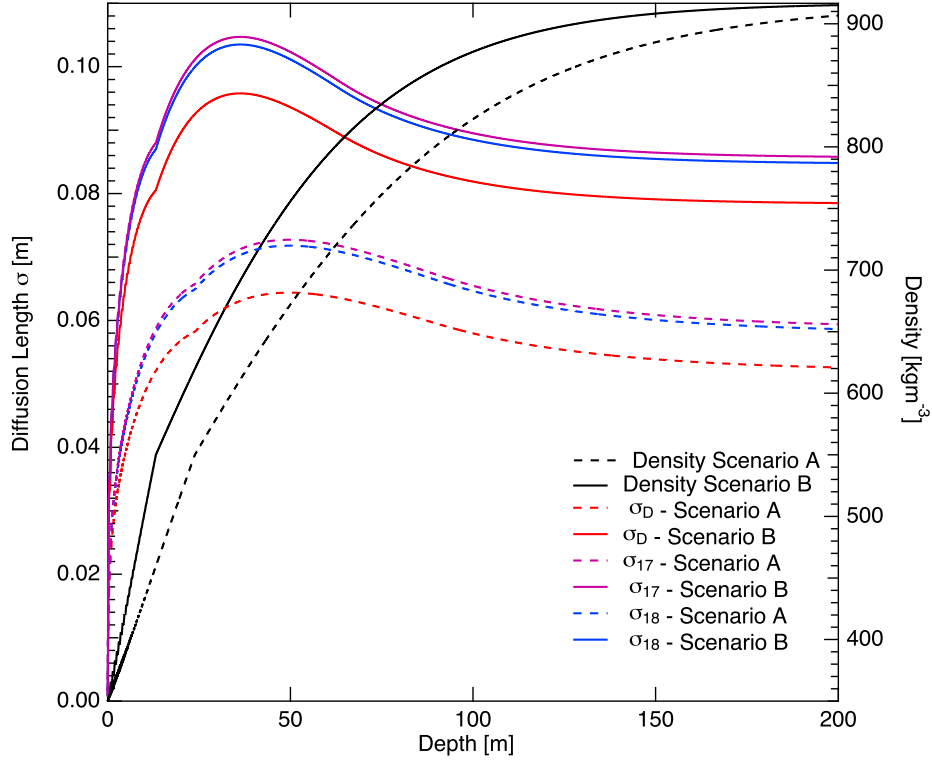}
			\caption{Diffusion length and density profiles (black) for case A (dashed lines: $T_{\mathrm{sur}} = -55 \,^{\circ} \mathrm{C}$,
				$A = 0.032 \;\mathrm{myr^{-1}}$) and B (solid curves: $T_{\mathrm{sur}} = -29 \,^{\circ} \mathrm{C}$,
				$A = 0.22 \;\mathrm{myr^{-1}}$).
				The increase in diffusion of the \delOx~(blue color), \delOxb~(purple color) and \delD~(red color) isotope signals are
				partially due to the compaction of the firn which moves the ice closer together. }
	\label{fig:diffusion_length_profiles}
		\end{center}

\end{figure}

\begin{figure}[]
	\vspace*{2mm}
	\begin{center}
		\includegraphics[width=0.7\textwidth]{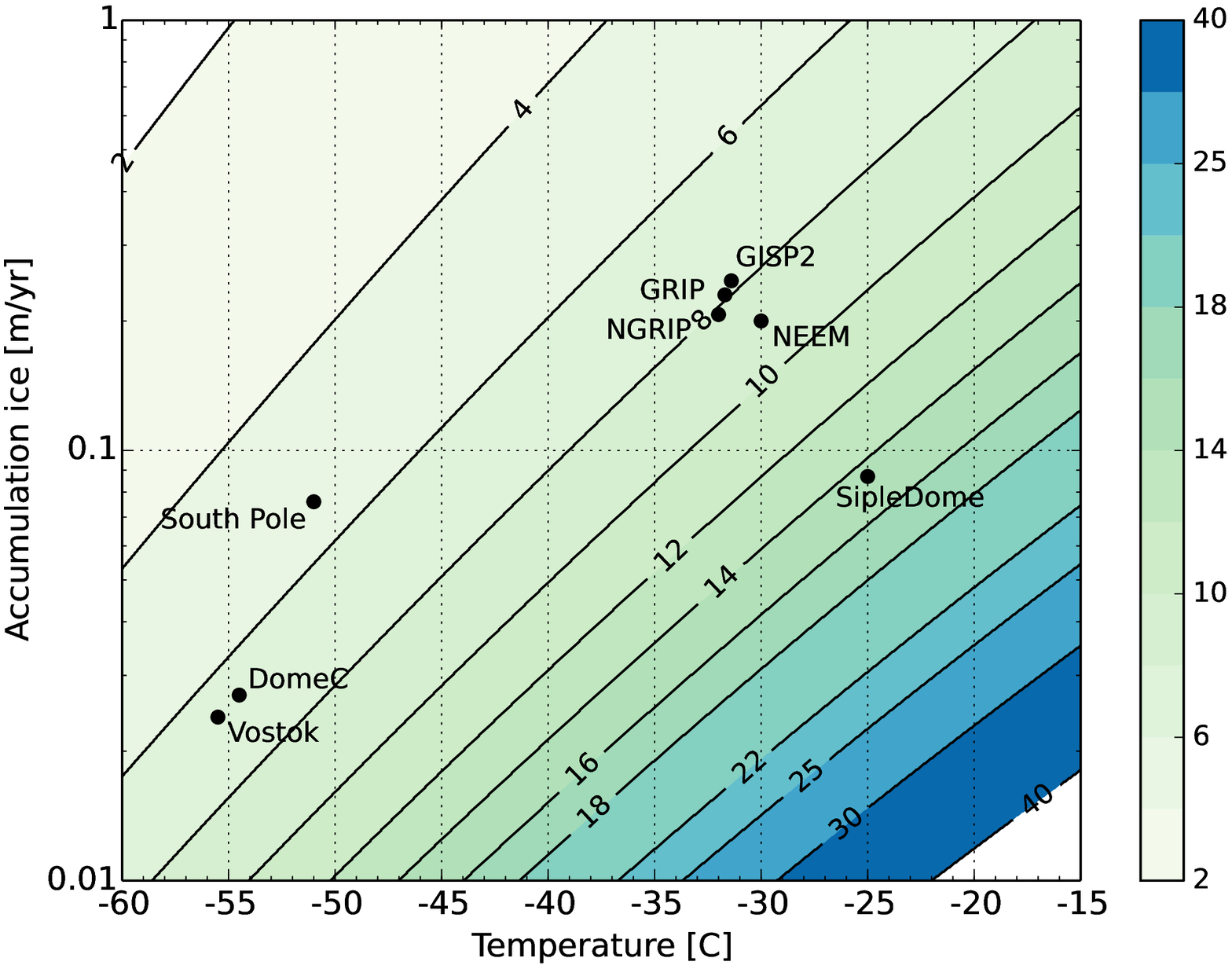}
		\caption{Modeled firn diffusion lengths [cm] for $\delta^{18}$O as a function of temperature and accumulation rate
				 (with $\rho_{\mathrm{co}}=804 \mathrm{kg m}^{-3}$ and
				$\rho_{\mathrm{o}}=330 \mathrm{kg m}^{-3}$) from \cite{Gkinis2014}.
				The contours indicate lines of constant diffusion length and the colorbar represents the diffusion length in cm.
				Here the combined impact of temperature and accumulation rate on the diffusion length is evident;
				while warm temperatures induce high diffusion lengths, a high accumulation rate reduces the diffusion length estimate.
				The firn diffusion lengths corresponding to a few ice core sites are plotted as a reference.}
		\label{fig:diffusion_map}
	\end{center}

\end{figure}

In Fig. \ref{fig:diffusion_length_profiles} we evaluate Eq. \ref{eq.difflengthintegration} for all three isotopic ratios
of water (\delOx, \delOxb, \delD) using boundary conditions characteristic of ice core sites from central Greenland
and the East Antarctic Ice Cap.
In Fig. \ref{fig:diffusion_length_profiles},
the transition between zone 1 and zone 2 densification (at the critical density $\rho_\mathrm{c} = 550\, \mathrm{kg m}^{-3}$) is
evident as a kink in both the densification and diffusion model.
For the first case we consider cold and dry conditions (case \textbf{A} hereafter)
representative of Antarctic ice coring sites
(e.g. Dome C, Vostok) with a surface temperature $T_{\mathrm{sur}} = -55 \,^{\circ} \mathrm{C}$ and
annual accumulation $A = 0.032 \;\mathrm{myr^{-1}\, ice\, eq.}$
For the second case we consider relatively warm and humid conditions (case \textbf{B} hereafter)
representative of central Greenlandic ice coring sites
(e.g. GISP2, GRIP, NorthGRIP) with a surface temperature $T_{\mathrm{sur}} = -29 \,^{\circ} \mathrm{C}$ and
annual accumulation $A = 0.22 \;\mathrm{myr^{-1}\, ice\, eq.}$
The general impact of surface temperature and accumulation rate on the firn diffusion length can be seen in Fig. \ref{fig:diffusion_map}.

\subsection{Isotope diffusion in the solid phase}
\label{ice_diffusion_section}
\begin{figure}[]
	\vspace*{2mm}
	\begin{center}
		\includegraphics[width=0.6\textwidth]{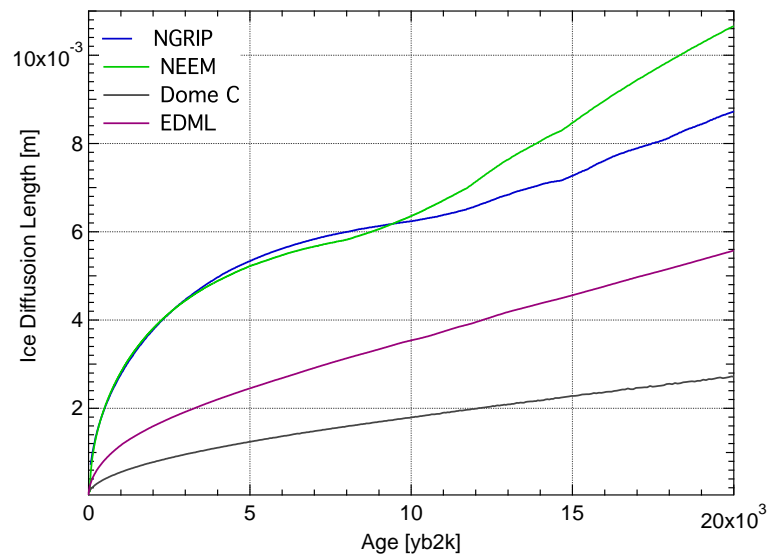}
		\caption{The ice diffusion length plotted with respect to age [b2k]
			for some selected sites from Greenland (NGRIP and NEEM) and Antarctica (Dome C and EDML).}
		\label{fig:ice_diffusion_multi}
	\end{center}
\end{figure}

Below the close-off depth, diffusion occurs in solid ice driven by the isotopic
gradients within the lattice of the ice crystals.
This process is several orders of magnitude slower than firn diffusion.
Several studies exist that deal with the estimate of the diffusivity
coefficient in ice \cite{Itagaki1964, Blicks1966, Delibaltas1966, Ramseier1967, Livingston1997}.
The differences resulting from the various diffusivity coefficients are small and
negligible for the case of our study (for a brief comparison between the different parameterizations,
the reader is referred to \cite{Gkinis2014}).
As done before by other similar firn diffusion studies \cite{Johnsen2000, Simonsen2011, Gkinis2014}
we make use of the parametrization given in \cite{Ramseier1967} as:
\begin{equation}
D_{\mathrm{ice}} = 9.2 \cdot 10^{-4} \cdot \exp \left(- \frac{7186}{T} \right) \mathrm{ m^2 s^{-1}} .
\label{eq.icediffusivity}
\end{equation}
Assuming that a depth--age scale as well as a thinning function are available for the
ice core a solution for the ice diffusion length is given by (SOM Sec. S3 for details):
\begin{equation}
\sigma_{\mathrm{ice}}^2 \!\left( t' \right) =
S\! \left( t' \right)^2 \,\int_0^{t'} 2 D_{\mathrm{ice}} \!\left( t \right) S \!\left( t \right) ^{-2} \,\mathrm{d} t.
\label{eq.icediffusion}
\end{equation}
In Fig. \ref{fig:ice_diffusion_multi} we have calculated ice diffusion lengths for four different cores (NGRIP, NEEM, Dome C, EDML).
For the calculation of $D_{\mathrm{ice}}$ we have used the borehole temperature profile of each core and assumed a steady state condition.
As the temperature of the ice increases closer to the bedrock, $\sigma_{\mathrm{ice}}$ increases nonlinearly
due to $D_{\mathrm{ice}}$ exponential temperature dependence.
When approaching these deeper parts of the core, the warmer ice temperatures enhance the effect of ice diffusion which then
becomes an important and progressively dominating factor in the calculations.
For the special case of the Dome C core (with a bottom age exceeding 800,000 years), $\sigma_{\mathrm{ice}}$
reaches values as high as 15 $\mathrm{cm}$.

\section{Reconstructing firn temperatures from ice core data}  \label{sec:methods}

Here we outline
the various temperature reconstruction techniques that can be employed for
paleotemperature reconstructions.
In order to avoid
significant overlap  with previously published works
e.g. \cite{Johnsen1977, Johnsen2000, Simonsen2011, Gkinis2014, vanderWel2015a}
we occasionally  point the reader to any of the latter
or/and refer to specific sections in the SOM.
We exemplify and illustrate the use of various techniques using synthetic data prepared such
that they resemble  two representative regimes of ice coring sites on the Greenland summit and the
East Antarctic Plateau. \\

\subsection{The single isotopologue diffusion}\label{sec:single_diffusion}

\begin{figure}[]
	\vspace*{2mm}
	\begin{center}
		\includegraphics[width=0.6\textwidth]{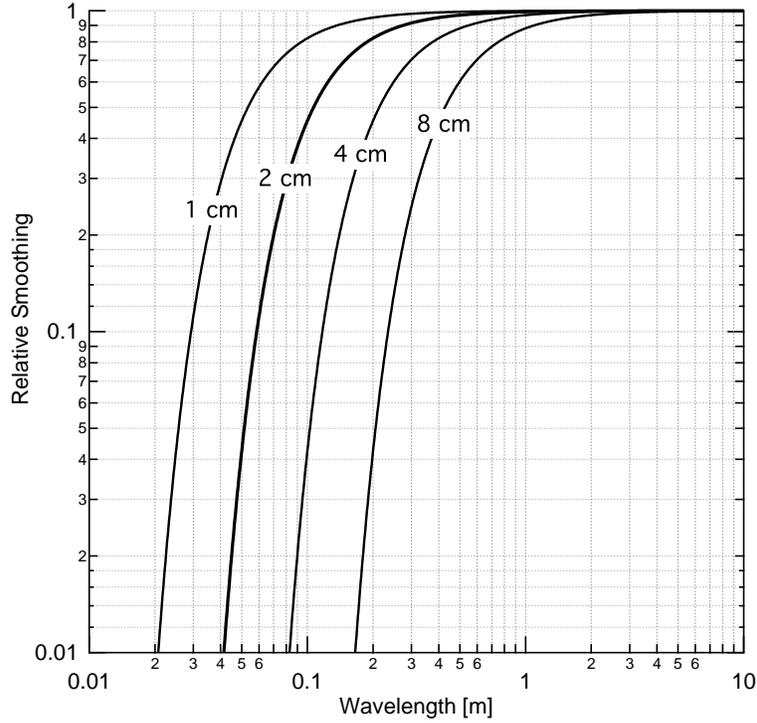}
		\caption{The smoothing effect of the diffusion transfer function demonstrated on
			a range of different wavelengths for $\sigma = 1,\, 2,\, 4\; \mathrm{and} \;8 \;	\mathrm{cm}$.}
		\label{fig:transfer_function}
	\end{center}
\end{figure}

As shown in Eq. \ref{eq.convolution}, the impact of the diffusion process can be mathematically
described as a convolution of the initial isotopic profile with a Gaussian filter.
A fundamental property of the convolution operation is that it is equivalent to  multiplication
in the frequency domain. The transfer function for the diffusion process will be given by the
Fourier transform of the Gaussian filter that will itself be a Gaussian function described by
\cite{Abramowitz1964, Gkinis2014}:
\begin{equation}
\label{ftrans_gaussian}
\mathfrak{F}
[ \mathcal{G} (z) ] =
\hat{\mathcal{G}} = {e}^{\frac{-k^2 \sigma^2}{2}} {} \enspace .
\end{equation}
In Eq. \ref{ftrans_gaussian}, $k = 2\pi f$ where $f$ is the frequency of the
isotopic time series.
In Fig. \ref{fig:transfer_function} we illustrate the effect of the diffusion transfer function on a range of wavelengths
for $\sigma = 1,\, 2,\, 4\; \mathrm{and} \;8 \;	\mathrm{cm.}$
Frequencies corresponding to wavelengths on the order of $50\,\mathrm{cm}$ and above remain largely unaltered while signals with
wavelengths shorter than $20\,\mathrm{cm}$ are heavily attenuated.

A data-based estimate of the diffusion length $\sigma$ can be obtained by looking at the power spectrum of the diffused
isotopic time series. Assuming a noise signal $\eta \left( k \right)$,
Eq. \ref{ftrans_gaussian} provides a model describing the power spectrum as:
\begin{equation}
P_s =    P_0(k) {e}^{-k^2 \sigma^2} + {\vert \hat{\eta} \left( k \right) \vert} ^{2}, \indent f \in \left[0,f_{\mathrm{Nq}}\right]
\label{eq:powerspectrum}
\end{equation}
where $f_{\mathrm{Nq}} = 1 /\left(2\Delta\right)$ is the Nyquist frequency that is defined by the sampling resolution $\Delta$.
$P_0(k)$ is the spectral density of the compressed profile without diffusion.
It is assumed independent of $k$ (now $P_0$) due to the strong depositional noise
encountered in high resolution $\delta$ ice core series \cite{Johnsen2000}.
Theoretically $|\hat{\eta}(k )|^2$ refers to white measurement noise.
As we will show later, real ice core data sometimes have a more red noise behavior.
A generalized model for the noise signal can be described well by autoregressive process of order 1 (AR-1).
Its power spectral density is defined as \cite{Kay1981}:
\begin{equation}
|\hat{\eta}(k )|^2 = \frac{\sigma_{\mathrm{\eta}}^2 \Delta}
 {\left| 1-a_1 \exp{\left( -i k  \Delta \right) } \right|^2} {},
\end{equation}
where $a_1$ is the AR-1 coefficient and $\sigma_{\mathrm{\eta}}^2$ is the variance of the noise signal.

\begin{figure}[]
	\vspace*{2mm}
	\begin{center}
		\includegraphics[width=0.65\textwidth]{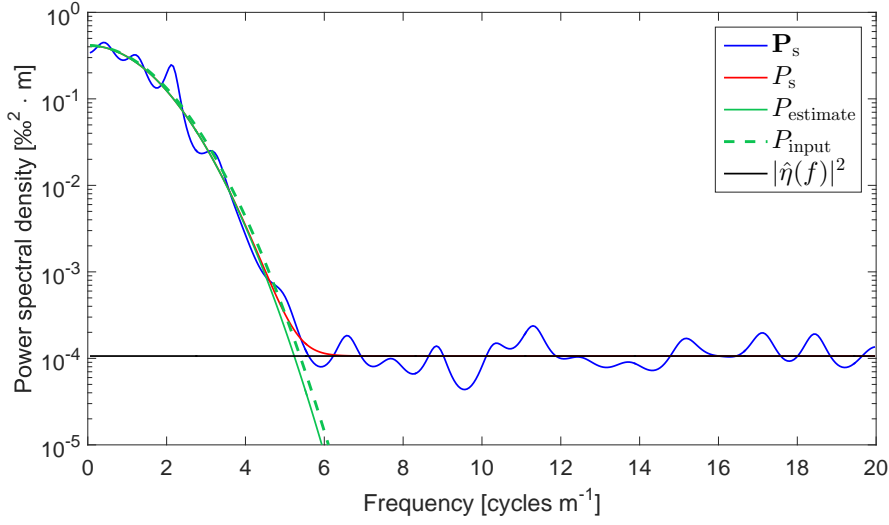}
		\caption{PSD of a synthetic \delOx~time series plotted with respect to frequency (blue curve).
				The red curve represents the complete model fit (Eq. \ref{eq:powerspectrum}).
				The green dashed curve represents the input diffusion
				and the the solid green curve represents the estimated diffusion length
				of the signal (uncorrected for sampling diffusion). The black curve represents the noise part of the fit.}
		\label{fig:synthetic_power_spectra}
	\end{center}
\end{figure}

In Fig. \ref{fig:synthetic_power_spectra}, an example of power spectra based on a synthetic time series is shown.
A description of how the synthetic time series is generated is provided in SOM Sec. S4.
The diffusion length used for the power spectrum in Fig. \ref{fig:synthetic_power_spectra} is equal to $8.50 \, \mathrm{cm}$.
The spectral estimate of the time series $\mathbb{P}_s$
is calculated using Burg's spectral estimation method
\cite{Kay1981} and specifically the algorithm presented in \cite{Andersen1974}.
Using a least--squares approach we optimize the fit of the model $P_s$ to the data-based $\mathbb{P}_s$ by varying the four
parameters $P_0$, $\sigma$, $a_1$ and $\sigma_{\mathrm{\eta}}^2$.
In the case of Fig. \ref{fig:synthetic_power_spectra},
the $\vert P_s - \mathbb{P}_s \vert^2$ least squares optimization resulted in
$P_0=0.32\,\permil^2\cdot \mathrm{m}$, $\sigma = 8.45 \, \mathrm{cm}$, $a_1 = 0.05$ and $\sigma_{\mathrm{\eta}}^2 =0.005 \, \permil^2$.

Assuming a diffusion length $\widehat{\sigma}_i^2$ is obtained for depth $z_i$ by means of
$\vert P_s - \mathbb{P}_s \vert^2$ minimization, one can calculate the equivalent diffusion
length at the bottom of the firn column $\sigma^2_{\mathrm{firn}}$ in order to estimate firn temperatures
by means of Eq. \ref{eq.difflengthintegration}.
In order to do this, one needs to take into account three necessary corrections
- (1) sampling diffusion, (2) ice diffusion and (3) thinning.
The first concerns the artifactually imposed
diffusion due to the sampling of the ice core.
In the case of a discrete sampling scheme with resolution $\Delta$ the additional
diffusion length is (SOM Sec. S5 for derivation):
\begin{equation}
\sigma^2_{\mathrm{dis}} = \frac{2\Delta^2}{\pi^2}\ln{\left(\frac{\pi}{2}\right)}.
\label{sampling_sigma}
\end{equation}
In the case of high resolution measurements carried out with CFA measurement systems, there
exist a number of ways to characterize the sampling diffusion length.
Typically the step or impulse response of the CFA system can be measured yielding a Gaussian
filter specific for the CFA system \cite{Gkinis2011, Maselli2013, Emanuelsson2015, Jones2017a}.
The Gaussian filter can be characterized by a diffusion
length $\sigma^2_{\mathrm{cfa}}$ that can be directly used to perform a sampling correction.
The second correction concerns the ice diffusion as described in Sec. \ref{ice_diffusion_section}.
The quantities $\sigma^2_{\mathrm{ice}}$ and $\sigma^2_{\mathrm{dis}}$ can be subtracted from
$\widehat{\sigma}_i^2$ yielding a scaled value   of $\sigma^2_{\mathrm{firn}}$ due to ice flow thinning.
As a result,  we can finally obtain the diffusion length estimate at the bottom of the firn column $\sigma^2_{\mathrm{firn}}$ (in meters of ice eq.):
\begin{equation}
\sigma^2_\text{firn} = \frac{\widehat{\sigma}_i^2 - \sigma^2_\text{dis} - \sigma^2_\text{ice}}{{\mathcal{S}(z)}^{2}}.
\label{eq.firn_data}
\end{equation}

Subsequently, a temperature estimate can be obtained by numerically finding the root of (for a known $A(z)$):
\begin{equation}
\label{eq:T_reconstruction}
\left(\frac{\rho_{co}}{\rho_i} \right)^2  \sigma^2(\rho = \rho_{\mathrm{co}},T(z), A(z)) - \sigma^2_\text{firn} = 0
\end{equation}
where $\sigma^2$ is the result of the integration in Eq. \ref{eq.difflengthintegration} from surface to
close--off density ($\rho_o  \rightarrow \rho_{\mathrm{co}}$). In this work we use a Newton-Raphson numerical scheme
\cite{RECIPES} for the calculation of the root of the equation.

The accuracy of the $\sigma^2_\text{firn}$ estimation and subsequently
of the temperature reconstruction
obtained based on it, depends on the three correction terms  $\sigma^2_{\mathrm{ice}}$,
$\sigma^2_{\mathrm{dis}}$ and the ice flow thinning $\mathcal{S}(z)$. For relatively shallow depths
where $\sigma^2_{\mathrm{ice}}$
is relatively small compared to $\widehat{\sigma}_i^2$, ice diffusion can be accounted for with
simple assumptions on the  borehole temperature profile and the ice flow.
In a similar way,
$\sigma^2_{\mathrm{dis}}$ is a well constrained parameter and depends only on the sampling
resolution $\Delta$ for discrete sampling schemes or the smoothing of the CFA measurement system.

Equation \ref{eq.firn_data} reveals an interesting property of the single isotopologue temperature estimation
technique. As seen, the result of the $\sigma^2_{\mathrm{firn}}$ calculation depends strongly on the
ice flow thinning quantity $\mathcal{S}(z)^2$. Possible errors in the estimation of $\mathcal{S}(z)^2$ due to imperfections
in the modelling of the ice flow will inevitably be propagated to the $\sigma^2_{\mathrm{firn}}$ value
thus biasing the temperature estimation. Even though this appears to be a disadvantage of the method,
in some instances, it can be a useful tool for assessing the accuracy of ice flow models.
Provided that for certain sections of the ice core there is a temperature estimate available based on
other reconstruction methods  (borehole thermometry, \delN / \delAr) it is possible to estimate
ice flow induced thinning of the ice core layers. Following this approach \cite{Gkinis2014} proposed
a correction in the existing accumulation rate history for the NorthGRIP ice core.

\subsection*{The annual spectral signal interference}
\begin{figure}[]
	\vspace*{2mm}
	\begin{center}
		\includegraphics[width=0.6\textwidth]{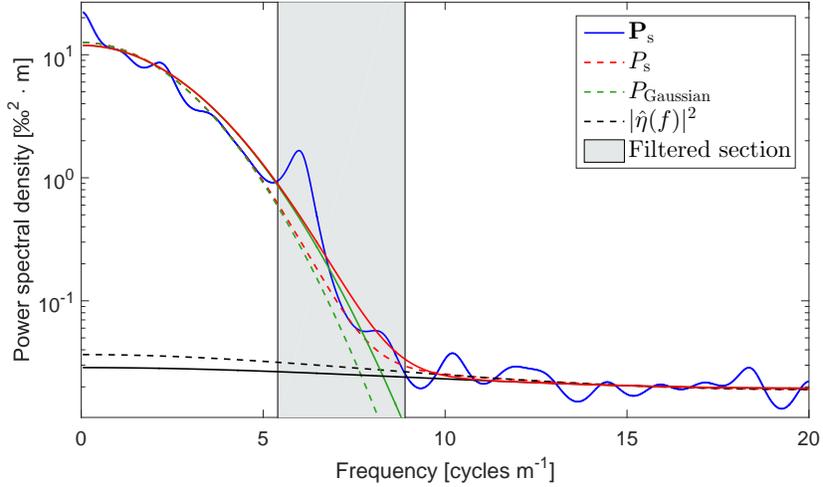}
		\caption{The interference of the annual spectral signal is seen
			in the PSD of the \delD~GRIP mid Holocene section. The regular fit is represented by the solid lines
			and the dashed lines represent the case where the weight function $w(f)$
			has filtered out this artifactual bias.} \label{fig:filter_function}
	\end{center}
\end{figure}

Depending on the ice core site temperature and accumulation conditions, spectral signatures of an annual isotopic signal can be apparent
as a peak located at the frequency range that corresponds to the annual layer thickness.
The resulting effect of such a spectral signature, is the artifactual biasing of the diffusion length estimation
towards lower values and thus colder temperatures.
Figure \ref{fig:filter_function} shows the PSD of the $\delta$D series for a mid Holocene section from the GRIP ice core
(drill site characteristics in Table \ref{tbl:drill_sites}).
A prominent spectral feature  is visible at  $f\approx6 \,\mathrm{cycles\,m^{-1}}$. This frequency is comparable to the
expected frequency of the annual signal at $6.1\,\mathrm{cycles\,m^{-1}}$ as estimated from the annual layer thickness reconstruction
of the GICC05 timescale \cite{Vinther2006}.

In order to evade the influence of the annual spectral signal on the diffusion length estimation, we propose the use of a weight
function $w(f)$ in the spectrum as:
\begin{equation}\label{eq:weight_function}
w(f)= \left\{
\begin{array}{ll}
0 &  f_{\lambda} - \mathrm{d}f_{\lambda} \leq f \leq
f_{\lambda} + \mathrm{d}f_{\lambda}\\
1                           &    f<f_{\lambda} - \mathrm{d}f_{\lambda}, f>f_{\lambda} + \mathrm{d}f_{\lambda}
\end{array} \right. {}
\end{equation}
where $f_\lambda$ is the frequency of the annual layer signal based on the reconstructed
annual layer thickness $\lambda$ and $\mathrm{d}f_{\lambda}$ is the range around the frequency $f_{\lambda}$
at which the annual signal is detectable.
The weight function is multiplied with the optimization norm $\vert P_s - \mathbb{P}_s \vert^2$.
Figure \ref{fig:filter_function} also illustrates the effect of the weight function on the estimation of $P_s$ and subsequently
the diffusion length value. When the weight function is used during the optimization process, there is an increase in the
diffusion length value by 0.3 cm, owing essentially to the exclusion of the annual signal peak from the minimization
of $\vert P_s - \mathbb{P}_s \vert^2$.
While the value of $f_{\lambda}$ can be roughly predicted, the value of $\mathrm{d}f_{\lambda}$ usually requires visual
inspection of the spectrum.

\subsection{The differential diffusion signal}\label{sec:diff_diffusion}
\begin{figure}[]
\vspace*{2mm}
\begin{center}
	\includegraphics[width=0.7\textwidth]{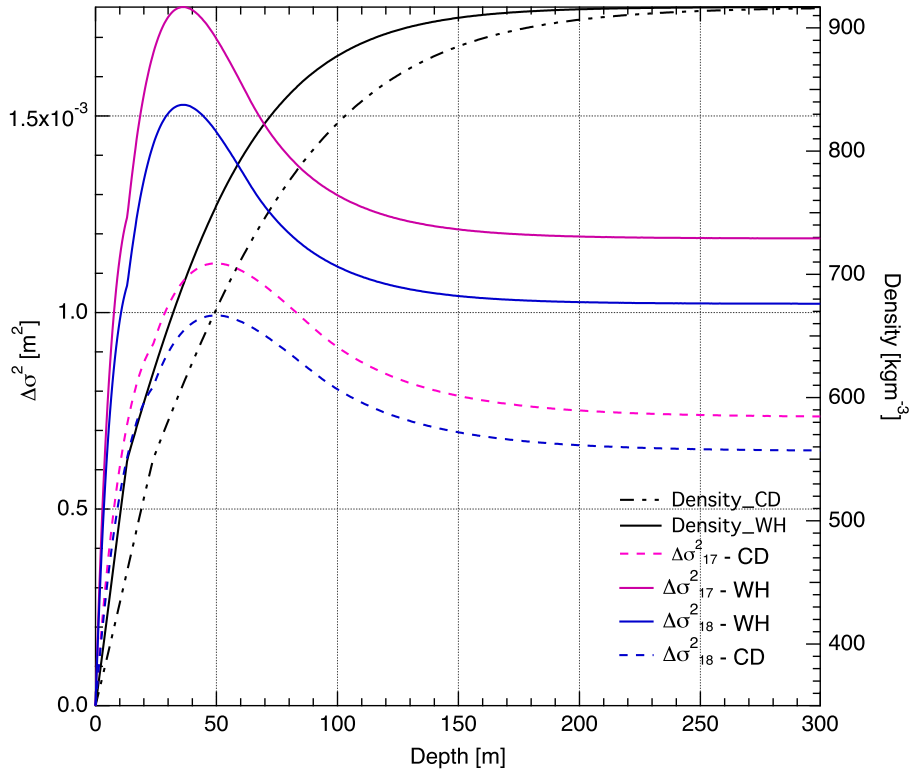}
	\caption{Differential diffusion length profiles for cases A (dashed lines) and B
		(solid lines) for ${}^{18}\Delta\sigma^2$  (blue) and ${}^{17}\Delta\sigma^2$ (purple).
		The density profiles are given in black.}
	\label{fig:diff_diffusion_example}
\end{center}
\end{figure}

A second-order temperature reconstruction technique is possible based on the differential signal between
\delOx~and \delD. Due to the difference in the fractionation factors and the air diffusivities between the oxygen and deuterium isotopologues,
a differential diffusion signal is created in the firn column. Based on the calculation of the diffusion lengths
presented in Fig. \ref{fig:diffusion_length_profiles} we then compute the differential diffusion lengths ${}^{17}\Delta\sigma^2$
and ${}^{18}\Delta\sigma^2$ where
\begin{equation}
{}^{17}\Delta\sigma^2 =  \sigma^2_{17} - \sigma^2_{\mathrm{D}} \mathrm{\,\,\,and\,\,}{}^{18}\Delta\sigma^2 =  \sigma^2_{18} - \sigma^2_{\mathrm{D}}.
\end{equation}
As it can be seen in Fig. \ref{fig:diff_diffusion_example} the differential diffusion length signal is slightly larger for the case
of ${}^{17}\Delta\sigma^2$ when compared to  ${}^{18}\Delta\sigma^2$.

One obvious complication of the differential diffusion technique is the requirement for dual measurements of the water
isotopologues, preferably performed on the same sample.
The evolution of IRMS techniques targeting the analysis of \delD~\cite{BIGELEISEN1952, Vaughn1998, Gehre1996, Begley1997} in ice cores has allowed for dual isotopic records
at high resolutions. With the advent of CRDS techniques and their customization for CFA measurements,
simultaneous high resolution measurements of both \delOx~and \delD~have become a routine procedure.

The case of \delOxb~is more complicated as the greater abundance of $^{13}$C than $^{17}$O rules out the
possibility for an IRMS measurement  at mass/charge ratio ($m/z$) of 45 or 29 using
CO$_2$ equilibration or reduction to CO respectively. Alternative approaches that exist include the
electrolysis method with CuSO$_4$ developed by  \cite{Meijer1998} as well as the fluorination method
presented by \cite{Baker2002} and implemented by \cite{Barkan2005} for dual-inlet IRMS systems.
These techniques target the measurement of the $^{17}${O}$_{\mathrm{excess}}$ parameter and are inferior for
\delOxb~measurements at high precision and have a very low sample throughput.
As a result, high resolution \delOxb~measurements from ice cores are currently non existent.
Recent innovations however in CRDS spectroscopy \cite{Steig2014} allow for simultaneous
triple isotopic measurements of \delD, \delOx~and \delOxb~in a way that  a precise and accurate measurement
for both \delOxb~and $^{17}${O}$_{\mathrm{excess}}$ is possible. Therefore high resolution ice core datasets of
\delD, \delOx~and \delOxb~should be expected in the near future.

\begin{figure}[]
\vspace*{2mm}
\begin{center}
	\includegraphics[width=0.8\textwidth]{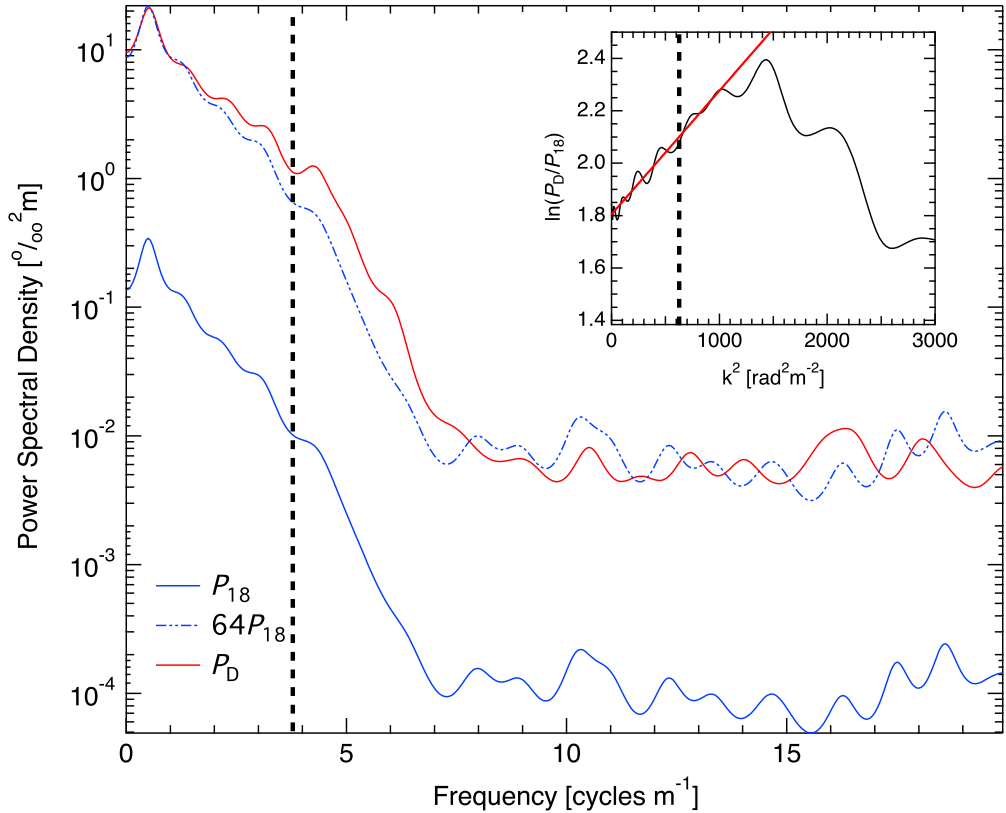}
	\caption{PSDs of synthetic \delOx~(blue) and \delD~(red) with respect to frequency where
		the inner subplot shows the $\ln \left({P_\mathrm{D}}/{P_{18}}\right)$ relation with respect to  $k^2$.
		The ${}^{18}\Delta\sigma^2$ value is determined from the slope of the linear fit in the subplot.
		The chosen cutoff frequency is marked by the vertical dashed line in both plots.}
	\label{fig:spectral_diff_diffusion_example}
\end{center}
\end{figure}

The following analysis is focused on the ${}^{18}\Delta\sigma^2$ signal but it applies equally to
the ${}^{17}\Delta\sigma^2$.
The stronger attenuation of the \delOx~signal with respect to the \delD~signal can be visually observed in the
power spectral densities of the two signals. As seen in Fig. \ref{fig:spectral_diff_diffusion_example} the $\mathbb{P}_{\mathrm{S18}}$
signal reaches the noise level at a lower frequency when compared to the $\mathbb{P}_{\mathrm{SD}}$ signal.
At low frequencies  with high signal to noise ratio we can calculate the logarithm of the ratio of the two power spectral densities as
(i.e. neglecting the noise term):
\begin{equation} \label{eq:diff_regression}
\ln \left(\frac{P_\mathrm{D}}{P_{18}}\right) \approx k^2\left( \sigma^2_{18} - \sigma^2_{\mathrm{D}}\right) +
\ln\left(\frac{P_{0_\mathrm{D}}}{P_{0_{18}}}\right) =
 {}^{18}\Delta \sigma^2\,k^2  + C.
\end{equation}
As seen in Eq. \ref{eq:diff_regression} and Fig. \ref{fig:spectral_diff_diffusion_example}
(synthetic generated \delOx~and \delD~data as in Sec. \ref{sec:single_diffusion}) an estimate of the
${}^{18}\Delta\sigma^2$ parameter can be obtained by a linear fit of
$\ln \left({P_\mathrm{D}}/{P_{18}}\right)$ in the low frequency area,
thus requiring only two parameters $({}^{18}\Delta\sigma^2\mathrm{ \,and\, } C)$ to be tuned.
An interesting aspect of the differential diffusion method, is that in contrast to the single isotopologue diffusion length,
${}^{18}\Delta\sigma^2_\mathrm{firn}$ is a quantity that is independent of the sampling and solid ice diffusion thus eliminating the
uncertainties associated with these two parameters. This can be seen by simply using Eq. \ref{eq.firn_data}:
\begin{equation}
{}^{18}\Delta\sigma^2_{\mathrm{firn}} = \frac{\hat{\sigma}_{18}^2 - \sigma_{\mathrm{dis}}^2 - \sigma_{\mathrm{ice}}^2}{\mathcal{S}(z)^2} -
\frac{\hat{\sigma}_\mathrm{D}^2 -
	\sigma_{\mathrm{dis}}^2 - \sigma_{\mathrm{ice}}^2}{\mathcal{S}(z)^2} =  \frac{\hat{\sigma}_{18}^2 -
	\hat{\sigma}_\mathrm{D}^2}{\mathcal{S}(z)^2}.
\end{equation}
Accurate estimates of the thinning function however still play a key role in the differential diffusion technique.
One more complication of the differential diffusion technique is the selection of the frequency range in
which one chooses to apply the linear regression. Often visual inspection is required in order to
designate a cut-off frequency until which the linear regression can be applied. In most cases
identifying the cut-off frequency, or at least a reasonable area around it is reasonably straight-forward.
Though in a small number of cases, spectral features in the low frequency area seem to have a strong
influence on the slope of the linear regression and thus on the ${}^{18}\Delta\sigma^2$. As a result,
visual inspection of the regression result is always advised in order to avoid biases.

Another  way of estimating the differential diffusion signal is to subtract the single diffusion
spectral estimates $\sigma^2_{18}$ and $\sigma^2_{\mathrm{D}}$.
Theoretically this approach should be inferior to the linear fit approach due to the fact that
more degrees of freedom are involved in the estimation of $\sigma^2_{18}$ and
$\sigma^2_{\mathrm{D}}$ (8 versus 2; 3 if the cutoff frequency is included).
Here we will test both approaches.

\subsection*{Linear correlation method}
\begin{figure}[]
	\vspace*{2mm}
	\begin{center}
		\includegraphics[width=0.7\textwidth]{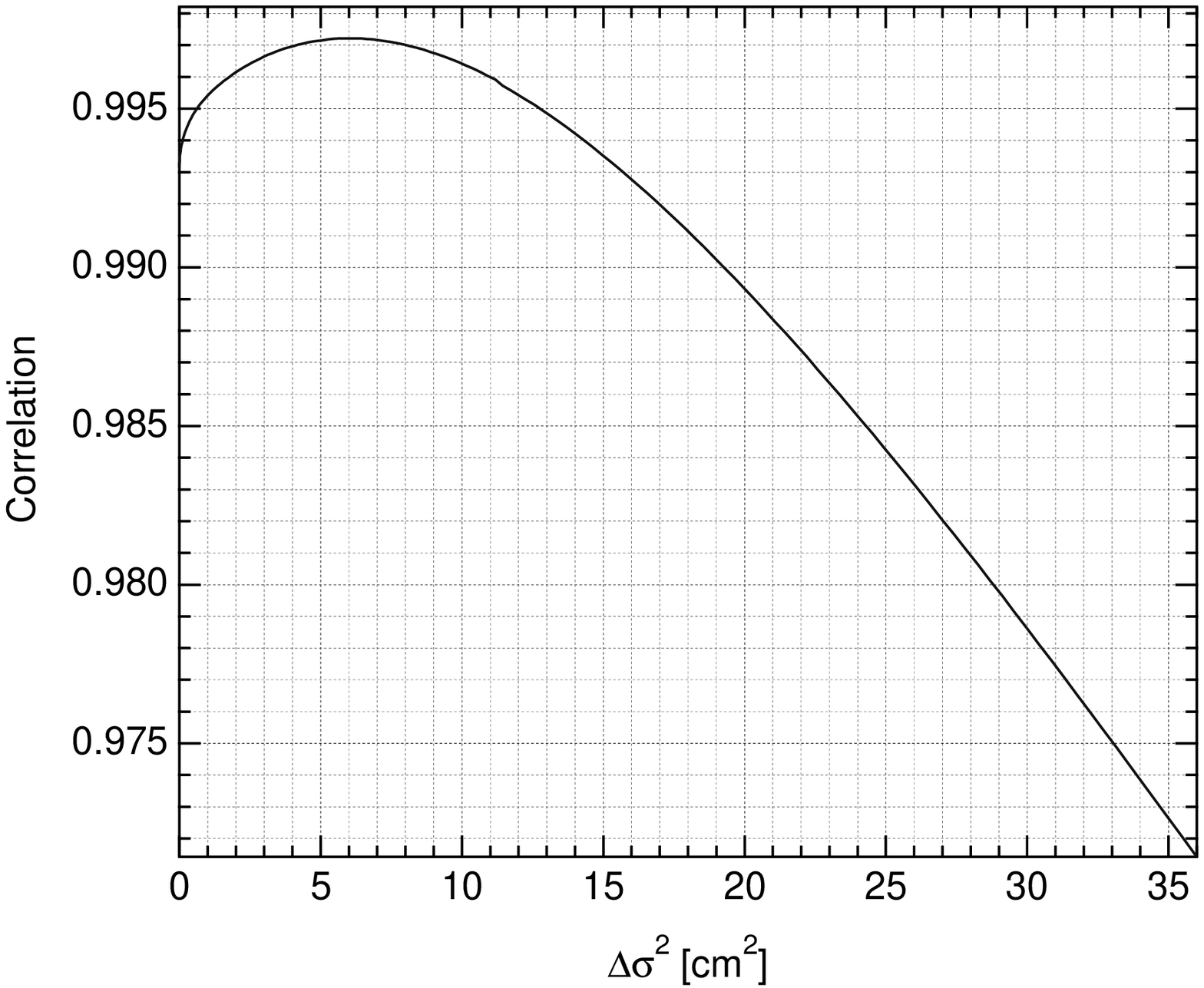}
		\caption{Correlation coefficient (\delcorrel) between the \delOx~and the forward--diffused \delD~series
		as a function of the estimated ${}^{18}\Delta\sigma^2$.
		The synthetic data represent a case A climate.}
		\label{fig:correlation_example}
	\end{center}
\end{figure}

An alternative way to calculate the differential diffusion signal ${}^{18}\Delta\sigma^2$
is based on the assumption that the initial precipitated isotopic signal presents a deuterium excess signal \Dxs~that
is invariable with time and as a consequence of this, the correlation signal between \delOx~and \delD~(hereafter \delcorrel) is expected
to have a maximum value at the time of deposition.
The \Dxs~signal is defined as the deviation from the metoric water line $\mathrm{d_{xs}}= \delta\mathrm{D}- 8\cdot\delta^{18}\mathrm{O} $ \cite{craig, Dansgaard}.
From the moment of deposition, the difference in diffusion between the \delOx~and \delD~signals results
in a decrease of the \delcorrel~ value. Hence, diffusing the \delD~signal
with a Gaussian kernel of standard deviation equal to ${}^{18}\Delta\sigma^2$ will maximize
the value of \delcorrel~\cite{vanderWel2015a} as shown in Fig. \ref{fig:correlation_example}.
Thus, the ${}^{18}\Delta\sigma^2$ value is found when the \delcorrel~value has its maximum.

This type of estimation is independent of spectral estimates of the \delOx~and \delD~time series
and does not pose any requirements for measurement noise characterization or selection of
cut-off frequencies. However uncertainties related to the densification and ice flow processes, affect
this method equally as they do for the spectrally based differential diffusion temperature estimation.
In this study, we test the applicability of the method on synthetic and real ice core data.
We acknowledge that the assumption that the \Dxs~signal is constant with time is
not entirely consistent with the fact that there is a small seasonal cycle in the \Dxs~signal \cite{Johnsen1989}.
It is thus likely to result in inaccuracies.

\subsection{The diffusion length ratio}

A third way of using the diffusion lengths as proxies for temperature can be based on the calculation
of the ratio of two different diffusion lengths. From Eq. \ref{eq.difflengthintegration} we can evaluate the ratio
of two different isotopologues $j$ and $k$ as:
\begin{equation}
\frac{\sigma_j^2 \left( \rho \right)}{\sigma_k^2 \left( \rho \right)} = \frac{\displaystyle \frac{\,1\,}{\rho^2} \int 2\rho^2
{\left( \frac{\mathrm{d}\rho}{\mathrm{d}t}\right)}^{-1}\! D_j \!\left( \rho \right) \,\mathrm{d}\rho}
{\displaystyle  \frac{\,1\,}{\rho^2} \displaystyle \int2\rho^2
{\left( \frac{\mathrm{d}\rho}{\mathrm{d}t}\right)}^{-1}\! D_k \!\left( \rho \right) \,\mathrm{d}\rho},
\label{eq.difflengthratio}
\end{equation}
and by substituting the firn diffusivities as defined in SOM Sec. S1 and according to
\cite{Johnsen2000} we get:
\begin{equation}
 \frac{\sigma_j^2 \left( \rho \right)}{\sigma_k^2 \left( \rho \right)} =
\frac{D_{\mathrm{a}j} \alpha_k}{D_{\mathrm{a}k} \alpha_j}\,
\frac{\displaystyle \frac{\,1\,}{\rho^2} \int 2\rho^2
{\left( \frac{\mathrm{d}\rho}{\mathrm{d}t}\right)}^{-1}\!  \frac{m\,p}{R\,T\,\tau}
\left(\frac{\,1\,}{\,\rho\,} - \frac{\,1\,}{\,\rho_{\mathrm{ice}}\,}\right) \,\mathrm{d}\rho}
{\displaystyle \frac{\,1\,}{\rho^2}  \int2\rho^2
{\left( \frac{\mathrm{d}\rho}{\mathrm{d}t}\right)}^{-1}\!  \frac{m\,p}{R\,T\tau}
\left(\frac{\,1\,}{\,\rho\,} - \frac{\,1\,}{\,\rho_{\mathrm{ice}}\,}\right) \,\mathrm{d}\rho} =
\frac{D_{\mathrm{a}j} \alpha_k}{D_{\mathrm{a}k} \alpha_j}.
\label{eq.difflengthratio_b}
\end{equation}
As a result, the ratio of the diffusion lengths is dependent on temperature through the parameterizations
of the fractionation factors ($\alpha$) and carries no dependence to parameters related to the densification rates nor the
atmospheric pressure. Additionally, it is a quantity that is independent of depth.
Here we give the analytical expressions of all the isotopologues combinations by substituting the diffusivities and the fractionation
factors:
\begin{align}
\sigma^2_{18}/\sigma^2_{\mathrm{D}} & = 0.93274\cdot \exp(16288/T^2-11.839/T)\\
\sigma^2_{17}/\sigma^2_{\mathrm{D}} & = 0.933\cdot \exp(16288/T^2-6.263/T)\\
\sigma^2_{18}/\sigma^2_{17} & = 0.99974\cdot \exp(-5.57617/T)
\label{eq.difflengthratio_c}
\end{align}

A data-based diffusion length ratio estimate can be obtained by estimating the single
diffusion length values as described in Sec. \ref{sec:single_diffusion}
and thereafter applying the necessary corrections as in Eq. \ref{eq.firn_data}.
An interesting aspect of the ratio estimation is that it is not dependent on the
ice flow thinning as seen below
\begin{equation}
\left( \frac{\sigma^2_{18}}{\sigma^2_\mathrm{D}} \right)_{\mathrm{firn}}  =
\frac{ \hat{\sigma}^2_{18} - \sigma^2_{\mathrm{dis}} - \sigma^2_{\mathrm{ice}} }{\hat{\sigma}^2_{\mathrm{D}} -
	\sigma^2_{\mathrm{dis}} - \sigma^2_{\mathrm{ice}} }.
\end{equation}
while the method still depends on the sampling and ice diffusion.
\begin{figure}[]
	\vspace*{2mm}
	\begin{center}
		\includegraphics[width=0.7\textwidth]{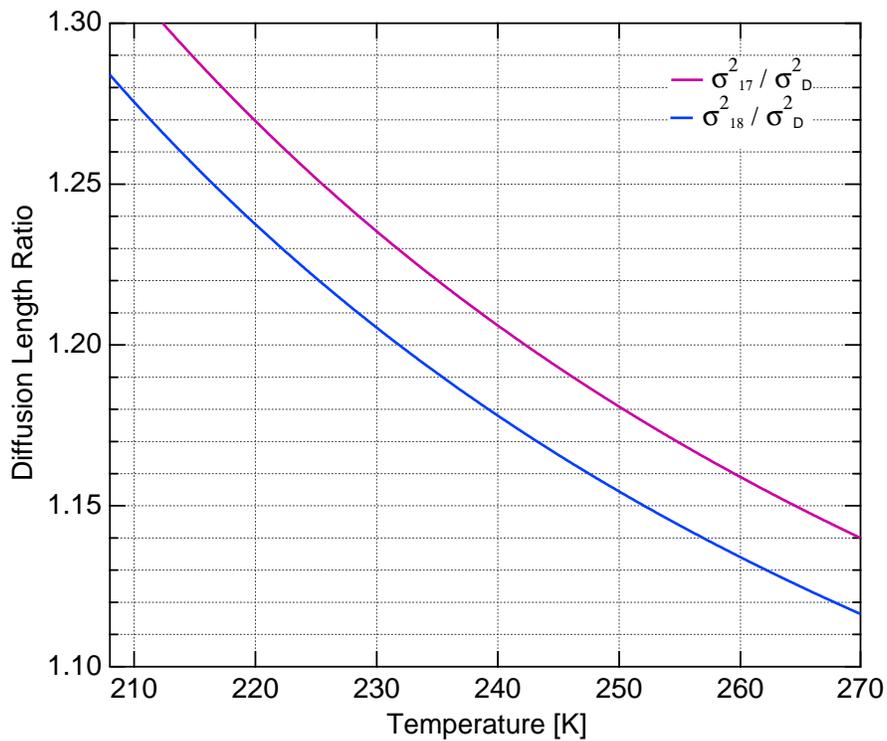}
		\caption{The diffusion length ratios $\sigma^2_{18}/\sigma^2_{\mathrm{D}}$ and $\sigma^2_{17}/\sigma^2_{\mathrm{D}}$
			with respect to temperature.
			The $\sigma^2_{18}/\sigma^2_{17}$ is almost constant at $0.975$ and omitted here due to its very low temperature sensitivity.}
		\label{fig:diff_len_ratio}
	\end{center}
\end{figure}


\section{Results}\label{sec:results}

\subsection{Synthetic data test} \label{sec:synthetic_results}
A first order test for the achievable accuracy and precision of the presented diffusion temperature reconstruction
techniques can be performed using synthetic isotopic data.
We generate synthetic time series of
\delOxb, \delOx~and \delD~using an AR-1 process and subsequently applying numerical diffusion with diffusion
lengths as calculated for case A and B (as presented in Fig. \ref{fig:diffusion_length_profiles}).
The time series are then sampled at a resolution of $2.5\,\mathrm{cm}$ and white measurement noise is added. Eventually,
estimates of diffusion lengths for all three isotopologues are obtained using the techniques we have described
in the previous sections.
A more detailed description of how the synthetic data are generated is outlined in SOM Sec. S4.

\begin{figure*}[]
	\vspace*{2mm}
	\begin{center}
		\includegraphics[width=\textwidth]{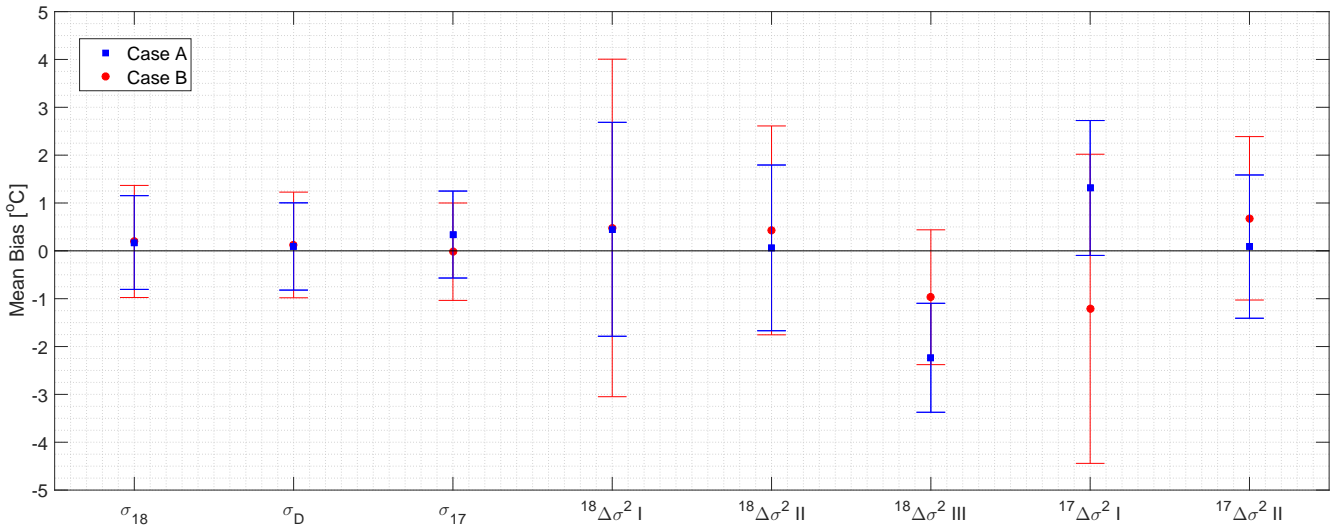}
		\caption{Mean biases for the single and differential diffusion techniques.
			The error bars represent 1 std of the estimated temperatures.}  \label{fig:synthetic_data_bias}
	\end{center}
\end{figure*}

The process of time series generation is repeated 500 times.
For each iteration, the quantities $\sigma_{17}$, $\sigma_{18}$,
$\sigma_\mathrm{D}$, ${}^{17}\Delta\sigma^2$, ${}^{18}\Delta\sigma^2$ and
the ratios $\sigma^2_{18}/\sigma^2_{\mathrm{D}}$,
$\sigma^2_{17}/\sigma^2_{\mathrm{D}}$ and $\sigma^2_{18}/\sigma^2_{17}$ are estimated.
The differential diffusion signals are estimated
using the three different techniques as described in Sec. \ref{sec:diff_diffusion}.
We designate the subtraction technique with
I, the linear regression with II and the correlation method with III.
For every value of the diffusion estimates we calculate a firn temperature where
the uncertainties related to the firn diffusion model ($A$, $\rho_{co}$, $\rho_o$, surface pressure $\mathrm{P}$, $\mathcal{S}$ and $\sigma_{ice}$ in Table \ref{tbl:model_unc}) are included.
For the total of the 500 iterations we calculate a mean firn temperature $\overline{T}$,
a standard deviation and a mean bias as:
\begin{equation}
\text{MB} = \frac{1}{N} \sum_{i = 1}^{N} T_i - T_{\text{sur}} = \overline{T} - T_{\text{sur}},
\end{equation}
where $i = 1,2,\dots,N$ signifies the iteration number, $T_i$ is the synthetic
data-based estimated temperature and $T_{\text{sur}}$ is the model forcing
surface temperature for the case A and B scenarios.
The results of the experiment are presented in Table \ref{tbl:synthetic_diff_lens} and the calculated
mean biases are illustrated in Fig. \ref{fig:synthetic_data_bias}.
The diffusion length ratio approach yields very large uncertainty bars (see Table \ref{tbl:synthetic_diff_lens}) and thus these results are not
included in Fig. \ref{fig:synthetic_data_bias}.

\begin{table}[H]
	\center
	\caption{The standard deviations of the input parameters.
		Most of the standard deviations are expressed as a percentage of the mean input value.}
	\label{tbl:model_unc}
	\begin{tabular}{c c c c c c c }
		\toprule
		Parameter & $A$  & $\rho_{co}$  & $\rho_o$  & $\mathrm{P}$  & $\mathcal{S}$ & $\sigma_{ice}$ \\
		\midrule
		Uncertainty &	$\pm 5 \% A_\mathrm{mean} $  &  $\pm20 \, \mathrm{kg m^{-3}}$  &  $\pm30 \, \mathrm{kg m^{-3}}$   &    $\pm 2 \% \mathrm{P}_\mathrm{mean} $  &     $\pm 1 \% \mathcal{S}_\mathrm{mean} $ 	 & $\pm 2 \% \sigma_{ice_\mathrm{mean}} $ \\
		\bottomrule
	\end{tabular}
\end{table}

\begin{table*}[]
	\center
	\resizebox{\columnwidth}{!}{%
	\begin{tabular}{r r r r r r r }
		\toprule
		& &  Case A & &  & Case B&\\
		\midrule
		&	Applied diffusion  &	Est. diffusion & Est. T [$^o\mathrm{C}$] &
		Applied diffusion	&	Est. diffusion 	& Est. T  [$^o\mathrm{C}$]\\
		\midrule
		$\sigma_{18}$	&	 $5.82$	 $\mathrm{cm}$&	$ 5.85\pm 0.14$ $\mathrm{cm}$	 &  $  -54.8 \pm 1.0 $ 	&	$8.50$  $\mathrm{cm}$&	$ 8.51 \pm 0.20$ $\mathrm{cm}$	& $ -28.8 \pm 1.2 $\\

		$\sigma_{\mathrm{D}}$&	$5.22$ $\mathrm{cm}$	&$ 5.23 \pm 0.12 $ $\mathrm{cm}$	 &	$-54.9  \pm 0.9 $ &	$ 7.86$	 $\mathrm{cm}$&	$7.86 \pm 0.18 $	$\mathrm{cm}$& 	$ -28.9\pm 1.1 $\\

		$\sigma_{17}$&	$5.90$ $\mathrm{cm}$	&$5.97 \pm 0.11 $ $\mathrm{cm}$	 &	$ -54.7 \pm 0.9$ &	$ 8.59$	 $\mathrm{cm}$&	$8.54 \pm 0.13 $	$\mathrm{cm}$& 	$-29.0 \pm 1.0 $\\

		${}^{18}\Delta\sigma^2$  I &$ 6.6	$ $\mathrm{cm^2}$&	$ 6.9\pm 1.1 $ $\mathrm{cm^2}$	& $ -54.6 \pm 2.2$   &	$10.3$  $\mathrm{cm^2}$	&	$10.7 \pm 2.0$ $\mathrm{cm^2}$	&$ -28.5 \pm 3.5 $\\

		${}^{18}\Delta\sigma^2$  II& $6.6$ $\mathrm{cm^2}$	&	$6.7 \pm 0.8$ $\mathrm{cm^2}$	& $ -54.9 \pm 1.7$	&	$ 10.3$ $\mathrm{cm^2}$	&$10.5 \pm 1.2 $	 $\mathrm{cm^2}$& $-28.6 \pm 2.2$\\

		${}^{18}\Delta\sigma^2$  III& $6.6$ $\mathrm{cm^2}$	&	$5.5 \pm 0.3$ $\mathrm{cm^2}$		&  $-57.2  \pm 1.1$	&	$10.3$	 $\mathrm{cm^2}$&	 $9.7 \pm 0.6 $  $\mathrm{cm^2}$& $-30.0 \pm 1.4 $\\

		${}^{17}\Delta\sigma^2$  I &$ 7.5	$ $\mathrm{cm^2}$&	$ 8.3 \pm 0.7 $ $\mathrm{cm^2}$	& $-53.7  \pm 1.4$   &	$12.0$  $\mathrm{cm^2}$	&	$ 12.2\pm 2.0 $ $\mathrm{cm^2}$	&$-30.2 \pm 3.2$\\

		${}^{17}\Delta\sigma^2$  II& $7.5$ $\mathrm{cm^2}$	&	$7.5 \pm 0.5$ $\mathrm{cm^2}$	& $-54.9  \pm 1.5 $	&	$12.0$ $\mathrm{cm^2}$	&$ 12.4\pm 1.0$	 $\mathrm{cm^2}$& $-28.3 \pm 1.7 $\\

		$\sigma^2_{18}/\sigma^2_{17} $&$	0.975$& $ 0.960\pm 0.027 $ 	& --------------	&$0.977$*	& 	$0.993 \pm 0.035$* 	& -------------- \\

		$\sigma^2_{18}/\sigma^2_\mathrm{D} $&$	1.24$& $1.25 \pm 0.04 $ 	& $-56.6 \pm 11.1$	&$1.17$	& 	$1.17 \pm 0.03$ 	& $-28.2 \pm 19.3$\\

		$\sigma^2_{17}/\sigma^2_\mathrm{D} $&$	1.28$& $ 1.31 \pm 0.03 $ 	& $  -62.8 \pm 7.0$	&$1.20$	& 	$1.18 \pm 0.04$ 	& $ -16.1 \pm 27 $\\
		\bottomrule
	\end{tabular}
}
	\caption{Simulations with synthetic data of a case A ($T_{\mathrm{sur}} = -55.0 \,^{\circ} \mathrm{C}$)
		and B ($T_{\mathrm{sur}} = -29.0 \,^{\circ} \mathrm{C}$).
		The diffusion lengths in the tabular are the firn diffusion lengths.
		Thus, this is before sampling, ice diffusion and thinning affected the input diffusion length.
		The estimated firn diffusion lengths are after correcting for sampling, ice diffusion and thinning (with their corresponding uncertainties).  }
	\label{tbl:synthetic_diff_lens}

\end{table*}

\subsection{Ice core data test} \label{sec:icecore_results}
We also use a number of high resolution, high precision ice core data, in order to benchmark the
diffusion temperature reconstruction techniques that we have  presented.
The aim of this benchmark test is to utilize the various reconstruction techniques for a range of
boundary conditions that is (a) as broad as possible with respect to mean annual surface temperature
and accumulation and (b) representative of existing polar ice core sites.
Additionally, we have made an effort in focusing on ice core  data sets that reflect conditions as close as possible to
present. As a result, the majority of the data sets presented here are from relatively shallow depths.
This serves a twofold purpose.
Firstly, it reduces the uncertainties regarding the ice flow that are considerably
larger for the deeper parts of the core.
Secondly, choosing to work with data sections as close to late Holocene
conditions as possible, allows for a comparison of the estimated temperature to the site's present temperature.
Although this is technically not a true comparison as the sites' surface temperatures have very likely
varied during the Holocene, we consider  it as a rough estimate of each techniques accuracy.
For those cases where it was not possible to obtain late Holocene isotopic time series, due to limited data availability,
we have used data originating from deeper sections of the ice cores with an age
of about 10ka b2k reflecting conditions of  the early Holocene. In Table \ref{tbl:drill_sites} we provide relevant
information for each data set as well as the present
temperature and accumulation conditions for each ice core site.
For five out of thirteen ice core data sets, we used a weight function of $w(f_{\lambda} - 0.5 \leq f \leq
f_{\lambda} + 3) = 0$ in order to remove the annual peak (see figures in SOM Sec. S6).

The data sets were produced using a variety of techniques
both with respect to the analysis itself (IRMS/CRDS), as well as with respect to the sample resolution and preparation
(discrete/CFA).
The majority of the data sets were analyzed using CRDS instrumentation. In particular the L1102i, L2120i and L2130i
variants of the Picarro CRDS analyzer were utilized for both discrete and CFA measurements of \delOx~and \delD.
The rest of the data sets were analyzed using IRMS techniques with either $\mathrm{CO}_2$ equilibration or
high temperature carbon reduction.
For the case of the NEEM early Holocene data set, we work with two data sections that span the same
depth interval and consist of discretely sampled and CFA measured data respectively.
Additionally, the Dome C and Dome F data sections represent conditions typical for the East Antarctic Plateau and are sampled
using a different approach ($2.5 \, \mathrm{cm}$ resolution discrete samples for the Dome C section and high resolution CFA measurements
for the Dome F section).

\begin{figure*}[]
	\vspace*{2mm}
	\begin{center}
		\includegraphics[width=\textwidth]{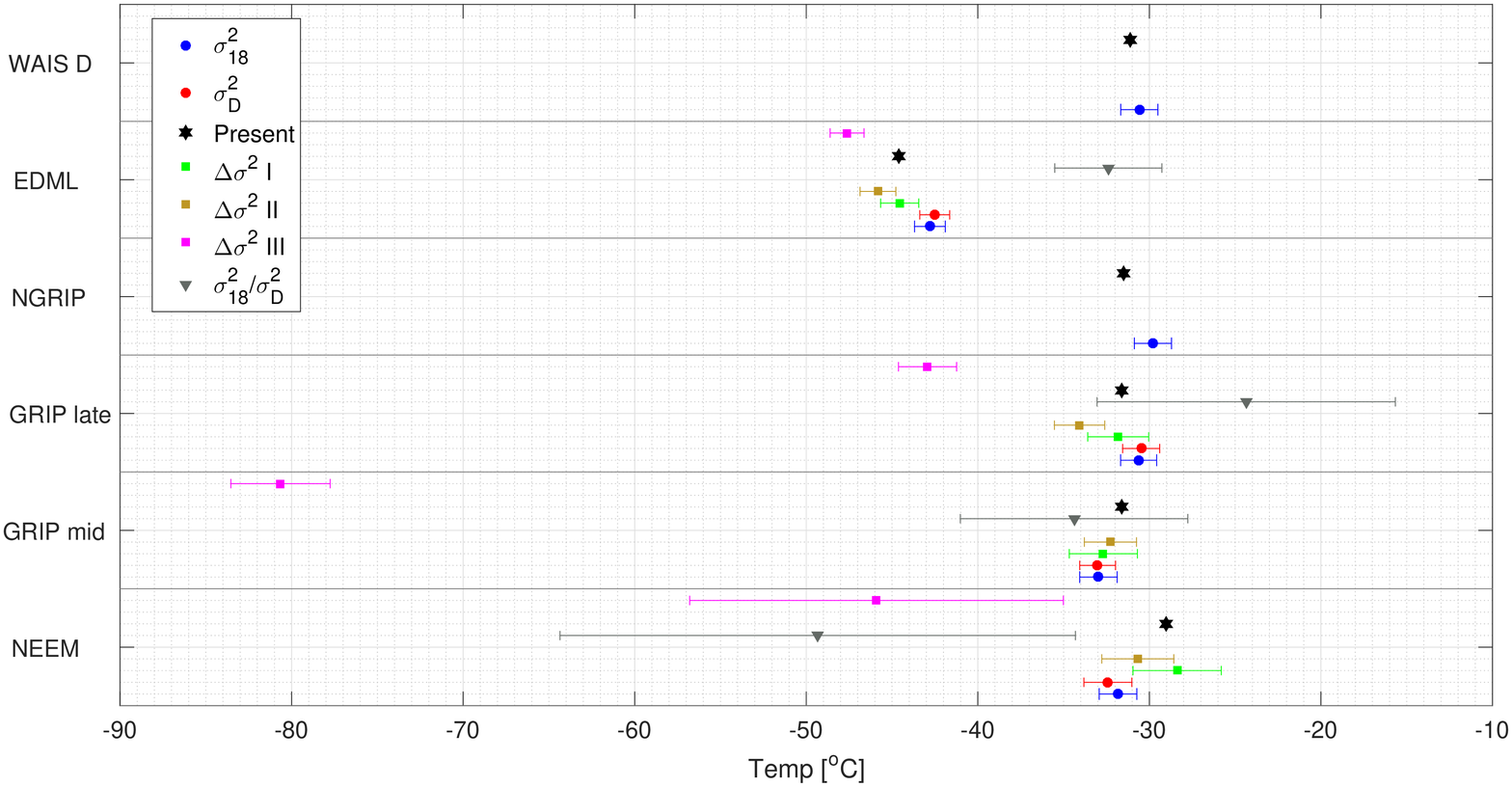}
		\caption{Late-mid Holocene section with reconstructed temperatures from the $\sigma^2_{18}$ (blue circles),
			$\sigma^2_{\mathrm{D}}$ (red circles), $\Delta\sigma^2$ I (green squares),
			$\Delta\sigma^2$ II (brown squares), $\Delta\sigma^2$ III (magenta squares) and
			${\sigma^2_{18}}/{\sigma^2_\mathrm{D}}$ (grey triangles) methods.
			The black stars represent the present annual mean temperatures at the sites.}  \label{fig:all_temps}
	\end{center}
\end{figure*}

\begin{figure*}[]
	\vspace*{2mm}
	\begin{center}
		\includegraphics[width=\textwidth]{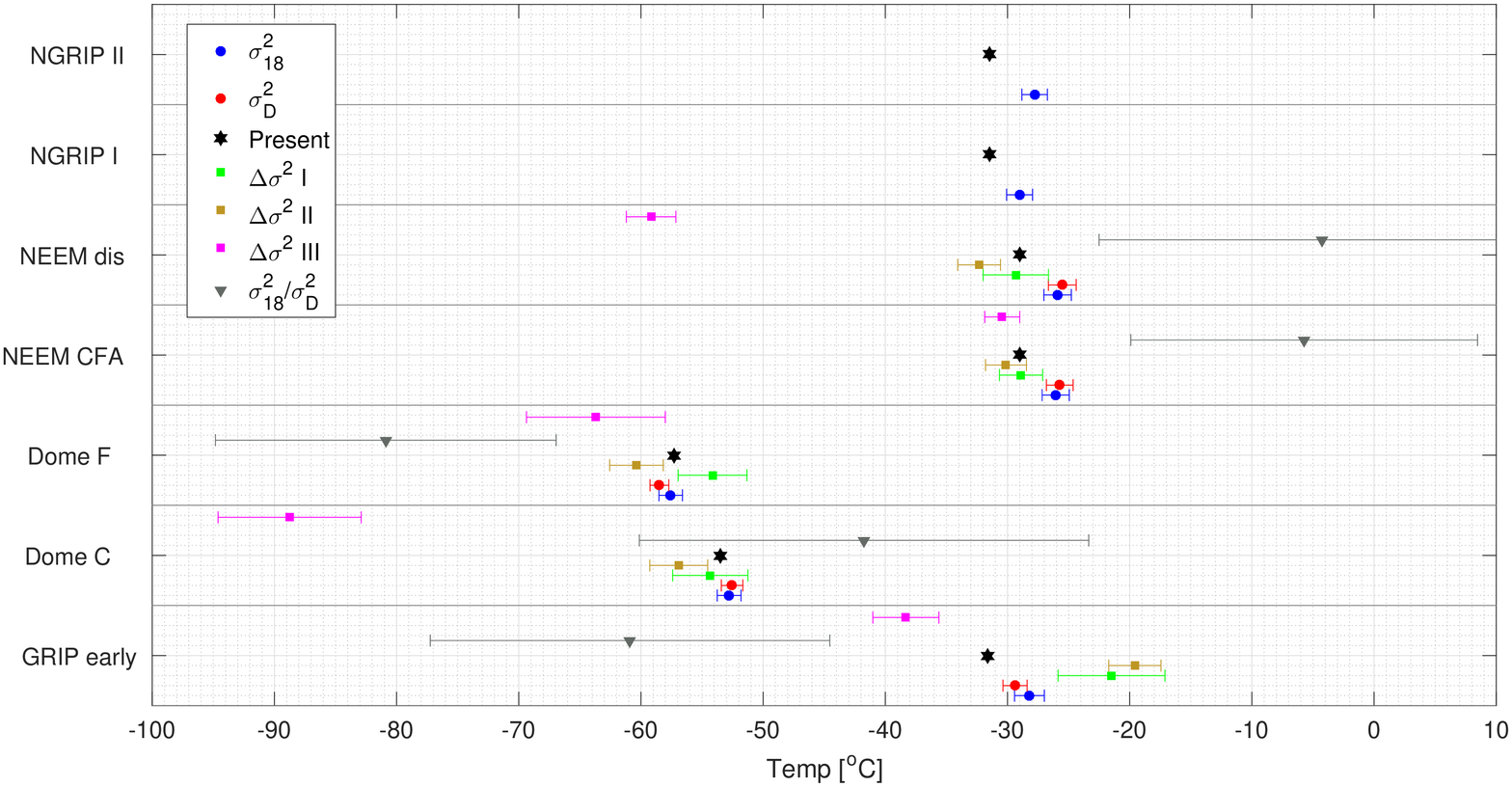}
		\caption{Early Holocene section with reconstructed temperatures from the $\sigma^2_{18}$ (blue circles),
			$\sigma^2_{\mathrm{D}}$ (red circles), $\Delta\sigma^2$ I (green squares),
			$\Delta\sigma^2$ II (brown squares), $\Delta\sigma^2$ III (magenta squares) and
			${\sigma^2_{18}}/{\sigma^2_\mathrm{D}}$ (grey triangles) methods.
			The black stars represent the present annual mean temperatures at the sites.}  \label{fig:early_temps}
	\end{center}
\end{figure*}

In a way similar to the synthetic data test, we apply the various reconstruction techniques on every ice core data
section. No reconstruction techniques involving \delOxb ~are presented here due to lack of
\delOxb ~data.
In order to achieve an uncertainty estimate for every reconstruction, we
perform a sensitivity test that is based on $N = 1000$ iterations.
Assuming that every ice core section
consists of $J$ \delOx ~and \delD ~points, then a repetition is based on a
data subsection with size $J'$ that varies in the interval $\left[ J/2, J \right]$.
This ``jittering'' of the subsection size happens around the midpoint of every section and $J'$ is
drawn from a uniform distribution.
Similar to the synthetic data tests, we also introduce uncertainties originating from the firn densification model, the ice flow  model
and ice diffusion (through the parameters: $A$, $\rho_{co}$, $\rho_o$, $\mathrm{P}$, $\mathcal{S}$ and $\sigma_{ice}$).
For every reconstruction method and every ice core site, we
calculate a mean and a standard deviation for the diffusion estimate, as well as a mean and a standard
deviation for the temperature.
Results are presented in Table \ref{tbl:icecore_results}.
The estimated temperatures for ice cores covering the late-mid Holocene and early Holocene are
shown in Fig. \ref{fig:all_temps} and \ref{fig:early_temps} respectively.

\subsection{The fractionation factors} \label{sec:results_fractionation}
We also test how the choice of the parameterization of the isotope fractionation factors ($\alpha_{18}$, $\alpha_\mathrm{D}$)
influences the reconstructed temperatures of ice core sections.
This is especially relevant for temperatures below $-40\mathrm{^oC}$, as the confidence of the
parameterized fractionation factors has been shown to be low for such cold temperatures \cite{Ellehoj2013}.
The low confidence is partly a consequence of two things a) it is difficult to avoid kinetic fractionation
in the measurement system and b) the water vapor pressure becomes small which makes it difficult to measure.
The experiments are typically performed with a vapor source with a known isotopic composition
that condenses out under controlled equilibrium conditions.
For temperatures below $-40\mathrm{^oC}$, single crystals have been observed growing against the flow of vapor
in the tubes and chambers of the experimental setup \cite{Ellehoj2013}.
This indicates that the water vapor experiences kinetic fractionation which disturbs the equilibrium process.
In order to avoid this, most models generally extrapolate the warmer experiments to cover colder temperatures.
Such extrapolations were performed in the parameterizations of \cite{Majoube1971} ($\alpha_{18}$) and \cite{Merlivat1967} ($\alpha_\mathrm{D}$)
which we used in the firn diffusivity parameterization (SOM Sec. S1).
Their experiments were conducted down to a minimum temperature of $-33\mathrm{^oC}$, and then extrapolated to colder temperatures.
Similarly, \cite{Ellehoj2013} estimated new values of $\alpha_{18}$ and $\alpha_\mathrm{D}$
by measuring in the range $-40\mathrm{^oC}$ to $0\mathrm{^oC}$.
Their results showed a $\alpha_\mathrm{D}$ parameterization that deviated significantly from that of \cite{Merlivat1967}.
A more recent study by \cite{Lamb2015} measured the value of $\alpha_\mathrm{D}$ in the range $-87\mathrm{^oC}$ to $-39\mathrm{^oC}$.
Their inferred equilibrium fractionation factors required a correction for kinetic effects.
By including such a correction and extrapolating to warmer temperatures,
they obtained a parameterization of $\alpha_\mathrm{D}$ with a slightly weaker temperature dependence than that of \cite{Merlivat1967}.
Moreover, their $\alpha_\mathrm{D}$ deviated significantly from the results of \cite{Ellehoj2013}.
Such discrepancies between the fractionation factor parameterizations underline the importance of addressing
how great an impact the potential inaccuracies have on the diffusion-based temperature proxy.

\begin{figure*}[]
	\vspace*{2mm}
	\begin{center}
		\includegraphics[width=\textwidth]{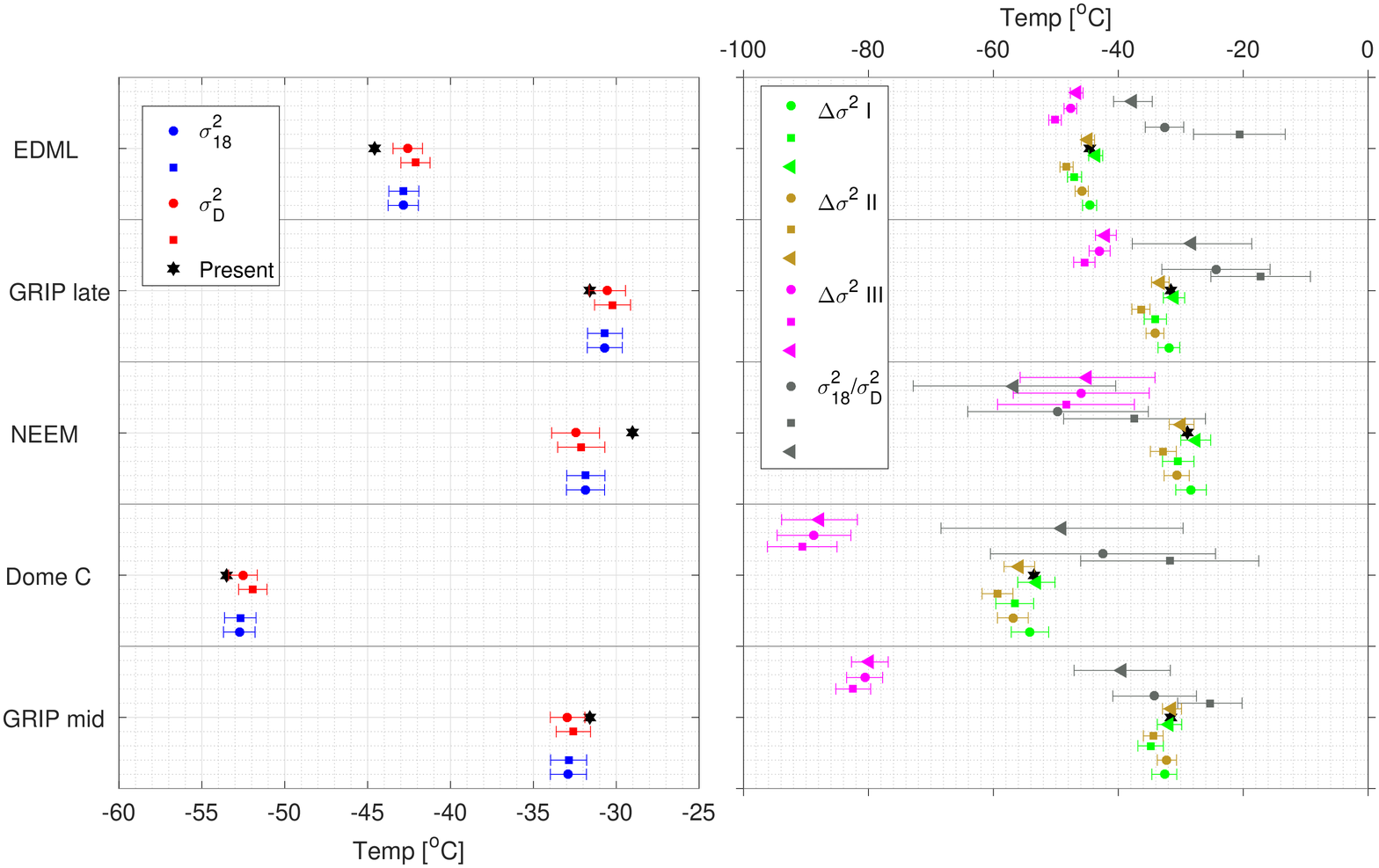}
		\caption{Temperature reconstructions based on
			different fractionation factor parameterizations.
			The left figure shows the single isotopologue methods and
			the right figure shows the dual isotope methods.
			Circles correspond to fractionation factors from \cite{Majoube1971,Merlivat1967},
			squares correspond to fractionation factors from \cite{Ellehoj2013} and  triangles from \cite{Lamb2015,Majoube1971}. }
		\label{fig:fractionation_temps}
	\end{center}
\end{figure*}

In this test, the procedure followed is common to that in Sec. \ref{sec:icecore_results} where a set of $N = 1000$ repetitions is performed
and both ``jittering'' of the data sets length and perturbation of input model variables takes place.
The results are displayed in Fig. \ref{fig:fractionation_temps}, where the temperatures resulting from the parameterizations of
\cite{Majoube1971} ($\alpha_{18}$) and \cite{Merlivat1967} ($\alpha_\mathrm{D}$) are compared to the temperatures resulting from the parameterizations
of \cite{Ellehoj2013} ($\alpha_{18}$, $\alpha_\mathrm{D}$) and \cite{Lamb2015} ($\alpha_\mathrm{D}$).
In the latter case, the parameterization of $\alpha_{18}$ from \cite{Majoube1971} is used for the dual diffusion length methods.

\begin{sidewaystable}[]
	\center
	\caption{Ice core data sections and the corresponding drill site characteristic. Sources of data: \cite{Steig2013}$^\mathrm{1}$,
		\cite{Oerter2004}$^\mathrm{2}$, \cite{Svensson2015}$^\mathrm{3}$, \cite{Gkinis2011b}$^\mathrm{4}$,\cite{Gkinis2011}$^\mathrm{5}$. Drill site characteristic sources:
		\cite{Bantaetal2008}$^\mathrm{a}$, \cite{Oerter2004,Veres2013}$^\mathrm{b}$, \cite{Watanabeetal2003,Kawamuraetal2003}$^\mathrm{c}$,
		\cite{Lorius1979}$^\mathrm{d}$, \cite{NGRIPmembers2004,Gkinis2014}$^\mathrm{e}$, \cite{Johnsen2000}$^\mathrm{f}$, \cite{Guillevic2013,Rasmussen2013}$^\mathrm{g}$.
	}\label{tbl:drill_sites}
	\begin{tabular}{l c c c c c c c c c}
		\toprule
		Site Sections &Depth & Age  & Present T  &
		A & P & Thinning& Meas. & Analysis&$\Delta$\\
		&[$\mathrm{m}$]&[$\mathrm{ka b2k}$]&[$^\mathrm{o}\mathrm{C}$]& [$\mathrm{m \, ice \, yr^{-1}}$]&
		[$\mathrm{Atm}$]& & &&[$\mathrm{cm}$]\\
		\midrule
		GRIP mid$^\mathrm{f}$&$753-776$ & 3.7 &$-31.6 $ & 0.23 & 0.65 & 0.71&$\delta$D, $\delta^{18}$O&2130&2.5\\
		GRIP late$^\mathrm{f}$&$514-531$  &2.4 &$-31.6 $ &0.23 & 0.65  &  0.79&$\delta$D, $\delta^{18}$O&2130&2.5\\
		WAIS 2005A$^\mathrm{a,1}$ &$120-150$  & 0.5 & $-31.1$ &0.22 & 0.77 & 0.97 &  	 $\delta^{18}$O&1102&5.0\\
		EDML$^\mathrm{b,2}$ & $123-178$&1.6&$-44.6$ & 0.08 & 0.67 & 0.93&$\delta$D, $\delta^{18}$O&IRMS&5.0\\
		NEEM$^\mathrm{g}$&$174-194 $&0.8 &$-29.0$ & 0.22 &  0.72 &  0.31&$\delta$D, $\delta^{18}$O&2120&2.5\\
		NGRIP$^\mathrm{e}$ &$174-194$ &0.9 &$-31.5$ & 0.20 & 0.67  & 0.49& $\delta^{18}$O&IRMS&2.5\\
		Dome F$^\mathrm{c,3}$& $302-307$& 9.6 & $-57.3$ &0.04 & 0.61  &  0.93 &$\delta$D, $\delta^{18}$O&CFA1102&0.5\\
		Dome C$^\mathrm{d,4}$ &$308-318$ & 9.9 &$-53.5 $ &0.04 & 0.65 &   0.93&$\delta$D, $\delta^{18}$O&IRMS&2.5\\
		GRIP early$^\mathrm{f}$&$1449-1466$ & 9.4 &$-31.6 $ & 0.23 & 0.65  &  0.42&$\delta$D, $\delta^{18}$O&2130&2.5\\
		NEEM dis$^\mathrm{g,5}$ &$1380-1392 $&10.9 &$-29.0$ & 0.22 &  0.72 &  0.31&$\delta$D, $\delta^{18}$O&2120&5.0\\
		NEEM CFA$^\mathrm{g,5}$& $1382-1399$& 10.9 &$-29.0 $& 0.22 & 0.72  & 0.31&$\delta$D, $\delta^{18}$O&CFA1102&0.5\\
		NGRIP I$^\mathrm{e}$&$1300-1320$ &9.1 &$-31.5$ & 0.18 & 0.67  & 0.55& $\delta^{18}$O&IRMS&5.0\\
		NGRIP II$^\mathrm{e}$ &$1300-1320$ & 9.1 &$-31.5$ & 0.18 & 0.67  & 0.55&$\delta^{18}$O&IRMS&5.0\\
		\bottomrule
	\end{tabular}
\end{sidewaystable}

\begin{table}[]
	\center

	\caption{Ice core results with the estimated firn diffusion lengths and their corresponding temperatures [$^\mathrm{o}$C].
		The units for the $\sigma_{18}$ and the $\sigma_{\mathrm{D}}$ values are expressed in $\mathrm{cm}$ and the unit for ${}^{18}\Delta\sigma^2$
		is expressed in $\mathrm{cm}^2$.  }\label{tbl:icecore_results}

	\resizebox{\columnwidth}{!}{%
	\begin{tabular}{l c c c c c c}
		\toprule
		Site Name & $\sigma_{18}$ & $\sigma_{\mathrm{D}}$ & ${}^{18}\Delta\sigma^2$  I & ${}^{18}\Delta\sigma^2$  II & ${}^{18}\Delta\sigma^2$  III & $\sigma^2_{18}/\sigma^2_{\mathrm{D}}$\\
		\midrule
		GRIP mid & 7.83 $\pm$ 0.17$\,\mathrm{cm}$& 7.20 $\pm$ 0.16$\,\mathrm{cm}$& 9.4 $\pm$ 1.0$\,\mathrm{cm}^2$& 9.6 $\pm$ 0.7$\,\mathrm{cm}^2$& 0.2$\pm$ 0.1$\,\mathrm{cm}^2$& 1.18$\pm$ 0.02\\
		& -33.0$\pm$ 1.1$\,^\mathrm{o}$C & -33.0$\pm$ 1.0$\,^\mathrm{o}$C& -32.7 $\pm$ 2.0$\,^\mathrm{o}$C& -32.3 $\pm$ 1.5$\,^\mathrm{o}$C& -80.6 $\pm$ 2.9$\,^\mathrm{o}$C& -34.4 $\pm$ 6.6$\,^\mathrm{o}$C\\

		GRIP late&8.52 $\pm$ 0.12$\,\mathrm{cm}$& 7.92 $\pm$ 0.16$\,\mathrm{cm}$& 9.9 $\pm$ 0.8$\,\mathrm{cm}^2$& 8.6 $\pm$ 0.5$\,\mathrm{cm}^2$& 4.8$\pm$ 0.5$\,\mathrm{cm}^2$& 1.16$\pm$ 0.02\\
		& -30.6 $\pm$ 1.1$\,^\mathrm{o}$C & -30.5 $\pm$ 1.1$\,^\mathrm{o}$C& -31.8 $\pm$ 1.8$\,^\mathrm{o}$C& -34.1 $\pm$ 1.5$\,^\mathrm{o}$C& -43.0 $\pm$ 1.7$\,^\mathrm{o}$C& -24.4 $\pm$ 8.7$\,^\mathrm{o}$C\\

		WAIS 2005A &7.05 $\pm$ 0.11$\,\mathrm{cm}$&--------------& --------------& --------------&--------------&--------------\\
		& -31.7 $\pm$ 1.1$\,^\mathrm{o}$C &-------------- &--------------&-------------- & --------------& --------------\\

		EDML &7.72 $\pm$ 0.09$\,\mathrm{cm}$& 7.12 $\pm$ 0.08$\,\mathrm{cm}$& 8.9 $\pm$ 0.3$\,\mathrm{cm}^2$& 8.1 $\pm$ 0.3$\,\mathrm{cm}^2$& 7.1$\pm$ 0.2$\,\mathrm{cm}^2$& 1.18$\pm$ 0.01\\
		& -42.8 $\pm$ 0.9$\,^\mathrm{o}$C & -42.5 $\pm$ 0.9$\,^\mathrm{o}$C& -44.6 $\pm$ 1.1$\,^\mathrm{o}$C& -45.9 $\pm$ 1.0$\,^\mathrm{o}$C& -47.6 $\pm$ 1.0$\,^\mathrm{o}$C& -32.4 $\pm$ 3.1$\,^\mathrm{o}$C\\

		NEEM&7.98 $\pm$ 0.22$\,\mathrm{cm}$& 7.20 $\pm$ 0.32$\,\mathrm{cm}$& 11.8 $\pm$ 1.6$\,\mathrm{cm}^2$& 10.2 $\pm$ 1.1$\,\mathrm{cm}^2$& 4.5$\pm$ 2.0$\,\mathrm{cm}^2$& 1.23$\pm$ 0.05\\
		& -31.8 $\pm$ 1.1$\,^\mathrm{o}$C & -32.4 $\pm$ 1.4$\,^\mathrm{o}$C& -28.4 $\pm$ 2.6$\,^\mathrm{o}$C& -30.7 $\pm$ 2.1$\,^\mathrm{o}$C& -45.9 $\pm$ 10.1$\,^\mathrm{o}$C& -49.3 $\pm$ 15.0$\,^\mathrm{o}$C\\

		NGRIP &9.24 $\pm$ 0.20 $\mathrm{cm}$&-------------- &--------------&-------------- & --------------&-------------- \\
		& -29.8 $\pm$ 1.1$\,^\mathrm{o}$C &--------------& --------------& --------------& --------------&--------------\\

		Dome F&5.76 $\pm$ 0.15$\,\mathrm{cm}$& 4.92 $\pm$ 0.06$\,\mathrm{cm}$& 9.0 $\pm$ 1.8$\,\mathrm{cm}^2$& 5.4 $\pm$ 0.8$\,\mathrm{cm}^2$& 4.4$\pm$ 1.9$\,\mathrm{cm}^2$& 1.37$\pm$ 0.08\\
		& -57.6 $\pm$ 1.0$\,^\mathrm{o}$C & -58.5 $\pm$ 0.8$\,^\mathrm{o}$C& -54.2 $\pm$ 2.8$\,^\mathrm{o}$C& -60.4 $\pm$ 2.2$\,^\mathrm{o}$C& -63.7 $\pm$ 5.7$\,^\mathrm{o}$C& -80.9 $\pm$ 14.0$\,^\mathrm{o}$C\\

		Dome C &6.97 $\pm$ 0.15$\,\mathrm{cm}$& 6.34 $\pm$ 0.08$\,\mathrm{cm}$& 8.4 $\pm$ 1.9$\,\mathrm{cm}^2$& 6.7 $\pm$ 1.1$\,\mathrm{cm}^2$& 0.4$\pm$ 0.4$\,\mathrm{cm}^2$& 1.21$\pm$ 0.05\\
		& -52.8 $\pm$ 1.0$\,^\mathrm{o}$C & -52.5 $\pm$ 0.9$\,^\mathrm{o}$C& -54.3 $\pm$ 3.0$\,^\mathrm{o}$C& -56.9 $\pm$ 2.3$\,^\mathrm{o}$C& -88.8 $\pm$ 5.9$\,^\mathrm{o}$C& -42.8 $\pm$ 18.4$\,^\mathrm{o}$C\\

		GRIP early &9.31 $\pm$ 0.24$\,\mathrm{cm}$& 8.25 $\pm$ 0.09$\,\mathrm{cm}$& 18.7 $\pm$ 4.0$\,\mathrm{cm}^2$& 20.4 $\pm$ 1.9$\,\mathrm{cm}^2$& 6.6$\pm$ 1.1$\,\mathrm{cm}^2$& 1.27$\pm$ 0.06\\
		& -28.2 $\pm$ 1.2$\,^\mathrm{o}$C & -29.4 $\pm$ 1.0$\,^\mathrm{o}$C& -21.5 $\pm$ 4.4$\,^\mathrm{o}$C& -19.6 $\pm$ 2.1$\,^\mathrm{o}$C& -38.4 $\pm$ 2.7$\,^\mathrm{o}$C& -60.9 $\pm$ 16.4$\,^\mathrm{o}$C\\

		NEEM dis &10.33 $\pm$ 0.19$\,\mathrm{cm}$& 9.72 $\pm$ 0.20$\,\mathrm{cm}$& 12.1 $\pm$ 1.8$\,\mathrm{cm}^2$& 10.0 $\pm$ 0.9$\,\mathrm{cm}^2$& 1.6$\pm$ 0.2$\,\mathrm{cm}^2$& 1.13$\pm$ 0.02\\
		& -25.9 $\pm$ 1.1$\,^\mathrm{o}$C & -25.5 $\pm$ 1.1$\,^\mathrm{o}$C& -29.3 $\pm$ 2.7$\,^\mathrm{o}$C& -32.3 $\pm$ 1.8$\,^\mathrm{o}$C& -59.2 $\pm$ 2.0$\,^\mathrm{o}$C& -4.2 $\pm$ 18.3$\,^\mathrm{o}$C\\

		NEEM CFA&10.27 $\pm$ 0.19$\,\mathrm{cm}$& 9.65 $\pm$ 0.18$\,\mathrm{cm}$& 12.3 $\pm$ 1.1$\,\mathrm{cm}^2$& 11.4 $\pm$ 0.9$\,\mathrm{cm}^2$& 11.2$\pm$ 0.6$\,\mathrm{cm}^2$& 1.13$\pm$ 0.01\\
		& -26.1 $\pm$ 1.1$\,^\mathrm{o}$C & -25.7 $\pm$ 1.1$\,^\mathrm{o}$C& -29.0 $\pm$ 1.8$\,^\mathrm{o}$C& -30.1 $\pm$ 1.7$\,^\mathrm{o}$C& -30.4 $\pm$ 1.4$\,^\mathrm{o}$C & -5.7 $\pm$ 14.2$\,^\mathrm{o}$C\\

		NGRIP I&9.68 $\pm$ 0.16$\,\mathrm{cm}$& --------------&--------------& --------------&--------------&--------------\\
		& -29.0 $\pm$ 1.1$\,^\mathrm{o}$C & --------------&--------------&-------------- &--------------& --------------\\

		NGRIP II &10.14 $\pm$ 0.17$\,\mathrm{cm}$&-------------- &--------------&--------------&-------------- & --------------\\
		& -27.8 $\pm$ 1.0$\,^\mathrm{o}$C & --------------&-------------- &-------------- &-------------- &--------------\\
		\bottomrule
	\end{tabular}
}
\end{table}

\section{Discussion}\label{sec:discussion}

\subsection{Synthetic data}\label{sec:discussion_synthetic}
Based on the results of the sensitivity experiment with
synthetic data, the following can be inferred.
Firstly, the three techniques based on the
single isotope diffusion, perform similarly and of all the techniques
tested, yield the highest precision with a $s_{\bar{T}} \approx 1.0 \,{}^{\circ} \text{C}$
(the average precision $s_{\bar{T}}$ of each technique is calculated
by averaging the variances of all simulations).
Additionally, the estimated temperatures $\overline{T}$ are within 1$s_{\bar{T}}$
of the forcing temperature  $T_{\text{sur}}$, a result pointing to a good performance with respect to
the accuracy of the temperature estimation.

The precision of the differential diffusion techniques is slightly
inferior to single diffusion with the subtraction
technique being the least precise of all three differential diffusion approaches ($s_{\bar{T}} \approx 2.6 \,{}^{\circ} \text{C}$).
A possible reason for this result may be the fact that the subtraction technique relies on the
tuning of 8 optimization parameters as described in Sec. \ref{sec:single_diffusion} and
\ref{sec:diff_diffusion}.
Both the linear fit and the correlation techniques yield precision estimates of $1.8 \,{}^{\circ} \text{C}$ and
 $1.3 \,{}^{\circ} \text{C}$, respectively.
Despite the high precision of the correlation technique, the tests shows that the technique has a bias toward colder temperatures.
The linear fit is therefore the most optimal of the differential diffusion techniques.
All 10 experiments utilizing differential diffusion methods, yield an accuracy that lies within the 2$s_{\bar{T}}$ range
(1$s_{\bar{T}}$ range for 9 out of 10 experiments). We can conclude that experiments involving the estimation
of the diffusion length ratio indicate that the latter are practically unusable due to very high uncertainties
with $s_{\bar{T}}$ averaging to a value of  $\approx 16 \,{}^{\circ} \text{C}$ for all four experiments.
A general trend that seems to be apparent for all the experiments, is that the results for the case A forcing yield
slightly lower uncertainties when compared to those for the case B forcing, likely indicating a
temperature and accumulation influence in the performance of all the reconstruction techniques.

\subsection{Ice core data}\label{sec:discussion_icecore}

\subsubsection{The estimation of diffusion length from spectra}

From the spectra presented in SOM Sec. S6,
we can see that the diffusion plus noise model (Eq. \ref{eq:powerspectrum}) provides good fits to the ice core data.
For ice core sections with a resolution equal to (or higher than) $2.5\,\mathrm{cm}$,
we start seeing a difference in the spectral signature of the noise tail between the data from Greenland and Antarctica.
The low accumulation Antarctic ice core sites seem to best represent the diffusion plus white noise model
used in the synthetic data test.
For instance, the PSD of Dome C in Fig. S32 resembles well that of the synthetic data
in Fig. \ref{fig:synthetic_power_spectra},
whereas a slightly more red noise tail is evident for the high accumulation sites on Greenland.
We don't know why the noise for some of the Greenlandic sites behaves differently,
but the white noise of the Antarctic ice core data coincides well
with isotopic signals that likely comprise of a few events per year and
is whiten due to post depositional effects such as snow relocation.
Nonetheless, the AR-1 noise model in Eq. \ref{eq:powerspectrum} describes both the red and white noise well.

An example of how sample resolution plays a role in assessing the value of the estimated diffusion length,
can be seen when visually comparing the spectra of the NEEM early Holocene data in Fig. S8 and S11.
The lack of sufficient resolution in Fig S11 (discrete $5\,\mathrm{cm}$ data) results in a poorly resolved noise signal.
On the contrary, the $0.5\,\mathrm{cm}$ resolution of the CFA obtained data (both datasets are from approximately the same depth interval)
allows for a much better insight into the noise characteristics of the isotopic time series and therefore a more robust diffusion length estimation.
Despite differences in the resolution of the power spectra,
the fitting procedure provides similar estimates of the firn diffusion lengths as seen in Table \ref{tbl:icecore_results}.
This result indicates that even though the diffusion length can be estimated with less certainty,
the diffusion length is still preserved in the signal which underlines how powerful a technique the spectral estimation of diffusion length is.

In this study, the annual peak is removed in five out of thirteen cases.
However, we do not see any distinguished multiannual variability manifested as spectral peaks.
A correction similar to that of the annual peak filter is therefore not implemented.
This does not necessarily mean that there is no imprint in those bands to start with,
but our analysis does not indicate this and these signals are either too weak to noticeably
affect the fits of the assumed model (i.e. diffusion plus noise)
or they cannot be resolved at all because their power lies lower than the measurement noise.

\subsubsection{The temperature reconstructions}
The precision $s_{\bar{T}}$ of each reconstruction technique has been quantified
by averaging the variances of the reconstructed temperatures (Table \ref{tbl:icecore_results}).
In accordance with the results from the synthetic data test, the most precise reconstructions are obtained when using the single isotope diffusion methods.
The single diffusion methods have a $s_{\bar{T}}$ of $1.1\,^\mathrm{o}\mathrm{C}$, while the differential diffusion methods ${}^{18}\Delta\sigma^2$ I, II and III
have a $s_{\bar{T}}$ of $2.6\,^\mathrm{o}\mathrm{C}$,
$1.9\,^\mathrm{o}\mathrm{C}$ and $4.8\,^\mathrm{o}\mathrm{C}$, respectively.
The correlation-based technique is hereby shown to be the least precise differential diffusion method.
This differs from the result of the synthetic data, where the correlation-based technique
had the most precise results.
Of the differential diffusion methods, the linear fit of the logarithmic ratio provides the most precise results,
with a precision similar to that found from the synthetic data (Sec. \ref{sec:discussion_synthetic}).
Of all the tested techniques, the diffusion length ratio method is the least precise with
a $s_{\bar{T}}$ of $11.8\,^\mathrm{o}\mathrm{C}$.
A similar precision was found from the synthetic data.

The perturbations of the model parameters help achieve a realistic view on the overall precision
and it facilitates a comparison between the single and the differential diffusion techniques.
Nonetheless, we want to emphasize that the presented precisions do not represent the absolute obtainable precision
of the diffusion-based temperature reconstruction techniques.
While the uncertainties presented in Table \ref{tbl:model_unc} represents typical Holocene values estimated
for Central Greenland and the East Antarctic Ice Cap, the input parameters' uncertainties in the firn diffusion model
are essentially both depth and site dependent.
For instance, we have a better knowledge about the ice flow thinning at a low accumulation site e.g. Dome C
compared to that of a high accumulation site e.g. NGRIP for early Holocene ice core data,
This is a result of the Dome C site's early Holocene period being at a depth of $300\,\mathrm{m}$ while the NGRIP site's
early Holocene period is at a depth of $1300\,\mathrm{m}$.
Additionally, it is more more difficult to estimate the glacial accumulation rate at sites where the
present day values already are very low.
Basically, inferring a change
between $3\,\mathrm{cm/yr}$ and $1.5\,\mathrm{cm/yr}$ (and how stable this $1.5\,\mathrm{cm/yr}$ estimate is during the glacial)
is much harder and with higher uncertainties compared to going from $23\,\mathrm{cm/yr}$ to $10\,\mathrm{cm/yr}$
(where annual layer thickness information is available from chemistry).
Similarly, $\rho_{co}$ and $\rho_o$ are better known for Holocene conditions and likely close to present day
values while glacial conditions represent a regime at which those values may change more considerably.
Thus, when utilizing the diffusion techniques on long ice core records,
we propose that the uncertainties of such model parameters and corrections should be
based on specific characteristics of the ice core site and the part (or depth) of the core under consideration.

It is not possible to quantify the accuracy of the methods when applied on short ice core data sections,
as the reconstructed temperatures represent the integrated firn column temperature.
Even though the firn diffusion model has a polythermal firn layer due to the seasonal temperature variation,
we can only estimate a single value of the diffusion length from the data
(the exact temperature gradients a layer has experienced is unknown).
The reconstructed temperatures should therefore not necessarily be completely identical to present day annual temperatures.
However, clear outliers can still be inferred from the data as Holocene temperature estimates that deviate with
$30\,^\mathrm{o}\mathrm{C}$ from the present day annual mean temperatures are unrealistic.

First we address the correlation-based and diffusion length ratio techniques as these two
methods result in temperatures that clearly deviate with present day annual mean temperatures (Fig. \ref{fig:all_temps} and \ref{fig:early_temps}).
Besides the low precision of the diffusion length ratio method,
temperature estimates using the the correlation-based and diffusion length ratio techniques
are highly inconsistent with the results of the
other techniques, with root-mean-square deviations (RMSD) varying from $21\,^\mathrm{o}\mathrm{C}$ to $34 \, ^\mathrm{o}\mathrm{C}$.
In addition, it can be seen that the correlation-based method results in significantly different
temperatures for the discretely and continuously measured NEEM section.
A similar difference is not found from the spectral-based methods. Instead, these provide
consistent temperatures independent of the processing scheme.
The generally poor performance of the correlation-based method on ice core data contradicts the high accuracy and precision of the synthetic reconstructions, and is most likely caused by an oversimplification of the relationship between $\delta\mathrm{D}$ and $\delta^{18}\mathrm{O}$. The generation of the synthetic data is based on the assumption that $\delta\mathrm{D} = 8\cdot \delta^{18}\mathrm{O} + 10 \permil$. However, this premise neglects the time dependent \Dxs~signal.
The correlation-based method can therefore be used to accurately reconstruct synthetic temperatures, while the accuracy and precision are much lower for ice core data,
as such data has been influenced by the \Dxs~signal.
In addition, these temperature estimates have been shown to be dependent on the sampling process. The correlation-based method therefore yields uncertain estimates of the differential diffusion length.

The temperature estimates originating from the $\sigma^2_{18}$ and $\sigma^2_{\mathrm{D}}$ methods are found to have a RMSD of $0.7\,^\mathrm{o}\mathrm{C}$. This shows that the $\sigma^2_{18}$ and $\sigma^2_{\mathrm{D}}$ methods result in similar temperatures, which is consistent with the high accuracies found from the synthetic data test.
Furthermore, the early Holocene ice core data from Greenland consistently shows reconstructed temperatures warmer than present day (Fig. \ref{fig:early_temps}),
which corresponds well with a HCO of around $3\,^\mathrm{o}\mathrm{C}$ warming as found by \cite{Dorthe1998, Vinther2009}.
With the exception of WAIS D, the estimated temperatures for the late-mid Holocene using the $\sigma^2_{18}$ and $\sigma^2_{\mathrm{D}}$ methods
are either slightly warmer or colder than present day (Fig. \ref{fig:all_temps}).
These sections represent ages ranging from $0.9$ to $3.7$ ka
and it is not unreasonable to assume that the sites' surface temperatures
have varied in time.
We emphasize that some of the presented ice core sections are as short as $15\,\mathrm{m}$,
and that such temperature estimates will potentially be more similar to present day when averaged over a long time series.

The temperature estimates of the ${}^{18}\Delta\sigma^2$ I method are similar to the present day annual temperature in six out of nine cases.
However, the results of the ${}^{18}\Delta\sigma^2$ I and II techniques have a RMSD of $3.8\,^\mathrm{o}\mathrm{C}$.
The seemingly accurate performance of the ${}^{18}\Delta\sigma^2$ I method could be either a coincidence or correct.
Two of the similar temperature results are from the NEEM early Holocene data that likely should have had warmer surface temperatures
than present day.
It is therefore difficult to select the most accurate results as both of the differential diffusion techniques
before performed well in the accuracy test with the synthetic data.
One should therefore not have a preferred technique without utilizing both methods on longer ice core sections.
Basically, the reconstructed temperatures could be similar when the temperatures have been averaged over a longer record.
Besides the internal differences in the results of the differential techniques, most of the temperature estimates do not
match the results of the single diffusion lengths.


\subsection{The fractionation factors} \label{sec:fractionation}
The temperature estimates resulting from the different fractionation factor parametrizations
are shown in Fig. \ref{fig:fractionation_temps}.
For each method, the influence of the choice of parametrization on the reconstructed
temperatures has been quantified by calculating the RMSD between temperature estimates of two parametrizations.
Comparing the parametrizations of \cite{Ellehoj2013} to those of \cite{Majoube1971} and \cite{Merlivat1967},
the RMSDs of reconstructions that are based on the single diffusion lengths $\sigma^2_{18}$ and $\sigma^2_{\mathrm{D}}$ are
$0.04^\mathrm{o}\mathrm{C}$ and $0.4^\mathrm{o}\mathrm{C}$.
Thus, it is evident that the choice of fractionation factors has an insignificant effect
on the results of the $\sigma^2_{18}$ method and a small effect on the results of the $\sigma^2_{\mathrm{D}}$ method.
The choice of parameterization has a greater effect on the temperatures of the ${}^{18}\Delta\sigma^2$ techniques,
where the temperature estimate of the ${}^{18}\Delta\sigma^2$ I, II and III techniques have RMSDs of
$2.3^\mathrm{o}\mathrm{C}$, $2.3^\mathrm{o}\mathrm{C}$ and $2.2^\mathrm{o}\mathrm{C}$, respectively.
Comparing the parametrization of \cite{Lamb2015} to that of \cite{Merlivat1967},
the temperatures of the ${}^{18}\Delta\sigma^2$ I, II and III techniques have RMSDs of
$0.9^\mathrm{o}\mathrm{C}$, $0.9^\mathrm{o}\mathrm{C}$ and $1.0^\mathrm{o}\mathrm{C}$, respectively.
In general, smaller RMSDs are found when comparing with temperature estimates based on the \cite{Lamb2015} parametrization.
For instance, comparing the temperatures of the ${\sigma^2_{18}}/{\sigma^2_\mathrm{D}}$ technique based on \cite{Lamb2015} with those of \cite{Merlivat1967},
the ${\sigma^2_{18}}/{\sigma^2_\mathrm{D}}$ technique yields a RMSD of $5.9^\mathrm{o}\mathrm{C}$,
while the RMSD is $11.0^\mathrm{o}\mathrm{C}$ when comparing the results based on the parametrizations of
\cite{Ellehoj2013} with those of \cite{Majoube1971} and \cite{Merlivat1967}.
There are two reasons to why the RMSDs are smaller when comparing with the \cite{Lamb2015} parametrization:
the parametrized $\alpha_\mathrm{D}$ of \cite{Merlivat1967} differs more with that of \cite{Ellehoj2013} than with that of \cite{Lamb2015},
and the same $\alpha_{18}$ parametrization is used when comparing with \cite{Lamb2015}.

The ${\sigma^2_{18}}/{\sigma^2_\mathrm{D}}$ method is significantly more influenced by the
fractionation factors.
The high RMSDs imply that even if the diffusion length ratio is estimated with high confidence,
the method is still too sensitive to the choice of parameterization.
This makes the method less suitable as a paleoclimatic thermometer.

\subsection{Outlook with respect to ice core measurements} \label{sec:outlook}
It is obvious from the analysis we present here that the type of isotopic analysis chosen has an
impact on the quality of the power spectral estimates and subsequently on the diffusion length estimation.
One such important property of the spectral estimation that is directly dependent on the nature of the isotopic analysis
is the achievable Nyquist frequency, defined by the sampling resolution $\Delta$ of the isotopic time series.
The value of the Nyquist frequency $f_{\mathrm{Nq}}$ sets the limit in the frequency space until which a power spectral estimate can be obtained.
The higher the value of  $f_{\mathrm{Nq}}$, the more likely it is that the noise part  ${\vert \hat{\eta} \left( k \right) \vert} ^{2}$
of the power spectrum will be resolved by the spectral estimation routine.
The deeper the section under study, the higher the required $f_{\mathrm{Nq}}$ due to the fact that the ice flow thinning
results in a progressively lower value for the diffusion length and as a result the diffusion part of the spectrum extents
more into the higher frequencies.
This effect manifests particularly in the case of the early Holocene Greenland sections of this study.
For the case of the NEEM early Holocene record, one can observe the clear benefit of the higher sampling resolution
by comparing the discrete ($\Delta = 5 \;\mathrm{cm}$) to the the CFA ($\Delta = 0.5 \;\mathrm{cm}$) data set.
Characterizing the noise  signal ${\vert \hat{\eta} \left( k \right) \vert} ^{2}$ is more straight forward in the case of the CFA data.
On the contrary, at these depths of the NEEM core, the resolution of $5 \;\mathrm{cm}$ results in the spectral estimation
not being able to resolve the noise signal.

The diffusion of the sampling and measurement process itself is a parameter that needs to be
thoroughly addressed particularly during the development and construction of a CFA system
as well as during the measurement of an ice core with such a system.
Ideally, one would aim for (a) a dispersive behavior  that resembles as close as possible that of Gaussian
mixing, (b) a measurement system diffusion length $\sigma_{\mathrm{cfa}}$ that is as low as possible
and (c) a diffusive behavior that is stable as a function of time.
Real measurements with CFA systems indicate that most likely due to surface effects in the experimental apparatus
that lead to sample memory, the transfer functions of such systems depart from the ideal model of
Gaussian dispersion showing a slightly skewed behavior.
For some systems, this behavior resembles more that of a slightly skewed Log-Normal distribution \cite{Gkinis2011, Maselli2013, Emanuelsson2015}
or a more skewed distribution that in the case of \cite{Jones2017a} requires the product of two Log-Normal distributions to be accurately modeled.
The result of this behavior to the power spectral density is still a matter of further study as high resolution datasets obtained with CFA systems
are relatively recent.

Additionally the accuracy of the depth registration is essential in order for accurate spectral estimates to be possible.
Instabilities in melt rates of the ice stick under consideration can in principle be addressed  and a first-order correction
can be available assuming a length encoder is installed in the system.
Such a correction though does not take into account the fact that due to the constant sample flow rate through the CFA system,
the constant mixing volume of the system's components (sample tubing, valves etc) will cause a variable mixing as melt rates change.
The magnitudes and importance of these variations are not easy to assess and more work will be required in the future in order
to characterize and correct for these effects.

Due to the recent advances in laser spectroscopy we expect measurements of the
$\delta^{17}$O signal to be a common output from analyzed ice cores.
As we showed with synthetic data, such a signal can also be used to reconstruct temperatures.
Especially the differential diffusion length of $\delta^{17}$O
and $\delta$D showed higher precision than that of $\delta^{18}$O and $\delta$D.
Such measurements however,
require that laboratories around the world have access
to well calibrated standards. Calibration protocols for $\delta^{17}$O have been suggested
\cite{Schoenemannn2013} although there is still a lack of $\delta^{17}$O values
for the International Atomic Energy Agency standards VSMOW
(Vienna Standard Mean Ocean Water) and SLAP (Standard Light Antarctic Precipitation).

\section{Conclusions}  
This study assessed the performance of six different diffusion-based temperature reconstruction
techniques.
By applying the methods on synthetic data, first order tests of accuracy and bias were demonstrated and evaluated.
Moreover, this approach facilitated precision estimates of each method.
The precision of each technique was further quantified by
utilizing every variety of the diffusion-based temperature proxy on thirteen high resolution
data sets from Greenland and Antarctica.
The results showed that the single diffusion length methods yielded similar temperatures
and that they are the most precise of all the presented reconstruction techniques.
The most precise of the three differential diffusion length techniques was the linear fit of
the logarithmic ratio.
The most uncertain way of reconstructing past temperatures was by employing the diffusion length ratio method.
The results from the correlation-based method were inconsistent to the results obtained through the spectral-based methods,
and the method was considered to yield uncertain estimates of the differential diffusion length.

It was furthermore shown that the choice of fractionation factor parametrization
only had a small impact on the results from the single diffusion length methods,
while the influence was slightly higher for the differential diffusion length methods.
The diffusion length ratio method was highly sensitive to the fractionation factor parametrization,
and the method is not suitable as a paleoclimatic thermometer.

In conclusion, despite that the dual diffusion techniques seem to be the more optimal choices
due to their independence of sampling and ice diffusion or densification and thinning processes,
the uncertain estimates should outweigh the theoretical advantages for Holocene ice core data.


\section*{acknowledgements}
The research leading to these results has received funding from the European Research Council under the
European Union's Seventh Framework Programme (FP7/2007-2013) grant agreement \#610055 as part
of the ice2ice project.
The authors acknowledge the support of the Danish National Research Foundation through
the Centre for Ice and Climate at the Niels Bohr Institute (Copenhagen, Denmark).
We would like to thank A. Schauer, S. Schoenemann, B. Markle and E. Steig for ongoing fruitful discussions
and inspiration through the years on all things related to water isotope analysis and modelling.
We thank our colleagues at Centre for Ice and Climate for their generous contribution,
especially those who have assisted in processing the ice cores.
We also thank the NEEM project for providing the NEEM ice core samples.
NEEM is directed and organized by the Center of Ice and Climate at the Niels Bohr Institute and US
NSF, Office of Polar Programs.
It is supported by funding agencies and institutions in Belgium (FNRS-CFB and
FWO), Canada (NRCan/GSC), China (CAS), Denmark (FIST), France (IPEV, CNRS/INSU, CEA and ANR),
Germany (AWI), Iceland (RannIs), Japan (NIPR), Korea (KOPRI), The Netherlands (NWO/ALW), Sweden (VR),
Switzerland (SNF), United Kingdom (NERC) and the USA (US NSF, Office of Polar Programs).

We would like to thank the three anonymous reviewers whose thoughtful comments helped improve and clarify the manuscript.

\end{document}